\documentclass[aps,twocolumn,raggedbottom,prb,showpacs,nobalancelastpage,amssymb,groupedaddress]{revtex4}
\usepackage{graphicx}
\usepackage{amsmath}
\newcommand{\bra}[1]{\langle #1 \lvert}
\newcommand{\ket}[1]{\lvert #1 \rangle}
\begin{document}
\title{Transport through a double-quantum-dot system with noncollinearly polarized leads}
\author{R.\, Hornberger, S.\, Koller, G.\, Begemann, A.\, Donarini, and M.\, Grifoni}
\affiliation{Institut f\"{u}r Theoretische Physik, Universit\"at
Regensburg, 93035 Regensburg, Germany}
\date{\today}
\begin{abstract}
We investigate linear and nonlinear transport in a double quantum
dot system weakly coupled to spin-polarized leads.
 In the linear regime, the
conductance as well as the nonequilibrium spin accumulation are
evaluated in analytic form. The conductance as a function of the
gate voltage exhibits four peaks of different height, with mirror
symmetry with respect to the charge neutrality point. As the
polarization angle is varied, due to exchange effects, the position and shape of the peaks
change in a characteristic way which preserves the electron-hole
symmetry of the problem. In the nonlinear regime various spin-blockade effects are observed. Moreover, negative
differential conductance features occur for noncollinear magnetizations
 of the leads. In the considered sequential
tunneling limit, the tunneling magnetoresistance (TMR) is always
positive with a characteristic gate voltage dependence for
noncollinear magnetization. If a magnetic field is added to the
system, the TMR can become negative.
\end{abstract}
 \maketitle

\section{Introduction}
Spin-polarized transport through nanostructures is attracting
increasing interest due to its potential application in
spintronics \cite{Maekawa02,NJP07} as well as in quantum computing
\cite{Awschalom02}. Downscaling magnetoelectronics devices to the
nanoscale implies that Coulomb interaction effects become
increasingly important \cite{Grabert92,Sohn97}. In particular, the
interplay between spin-polarization and Coulomb blockade can give
rise to a complex transport behavior in which both the spin and
the charge of the ''information carrying" electron play a role.
 This has been widely demonstrated by many experimental
 studies  on  single-electron transistors (SETs) with ferromagnetic
leads, with  central element being either a ferromagnetic particle
\cite{Ono97,Schelp97,Yakushiji05}, normal metal particles
\cite{Zhang05,Bernand-Mantel06}, a two-level artificial molecule
\cite{Pioro-Ladriere03},  a  $C_{60}$ molecule \cite{Pasupathy04},
or a carbon nanotube \cite{Sahoo05}, showing the increasing
complexity and variety of the investigated systems. Initially, the
theoretical work was mainly focused on  the difference in the
transport properties for parallel or antiparallel magnetizations
in generic
 spin-valve SETs
\cite{Barnas98,Takahashi98,Majumdar98, Korotkov99,
Brataas99,BrataasWang01,WeymannKönig05,Gorelik05,CottetChoi06,Fransson06b,Weymann07}.
More recently, the interplay between spin and interaction effects
for noncollinear magnetization configurations has attracted quite
some interest both in systems with a continuous energy spectrum
\cite{Balents01,Bena02,Pedersen05,Wetzels05}, as well as in
single-level quantum dots
\cite{Sergueev02,König03,Braig05,Rudzinski05,Fransson05,Weymann05,Mu06,WeymannBarnas07},
many-level nanomagnets \cite{Parcollet06} and in carbon nanotubes
quantum dots \cite{Koller07}.
In the noncollinear case, a much richer physics is expected than
in the collinear one. For example, two separate exchange effects
have to be taken into account. On the one hand, there is the
non-local interface exchange, in scattering theory for
non-interacting systems described by the imaginary part of the
spin-mixing conductance \cite{Brataas}, and which in the context
of current-induced magnetization dynamics acts as an  effective
field \cite{Stiles}. Such an effective field has been found
experimentally to strongly affect the transport dynamics in spin
valves with MgO tunnel junctions \cite{Tulapurkar}. This effect
has recently also been involved to explain negative tunneling
magnetoresistance effects in carbon nanotube spin valves
\cite{Sahoo05} and called spin-dependent interface phase shifts
\cite{Cottet06,CottetChoi06}. The second exchange term is an
interaction-dependent exchange effect due to virtual tunneling
processes that is absent in non-interacting systems
\cite{Balents01,König03,Wetzels05,Koller07}.  This latter exchange
effect is potentially attractive for quantum information
processing, since it allows to switch on and off magnetic fields
in arbitrary directions just by a gate electric potential.
%
\begin{figure}
\includegraphics[height=2.4cm,width=5cm]{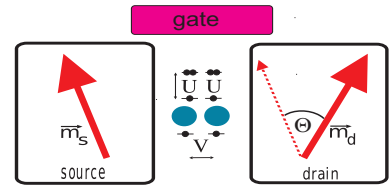}
 \caption{(Color online) Schematic picture of the model: A double quantum dot
system attached to polarized leads. The significance of the
on-site and inter-site interactions $U$ and $V$, respectively, is
depicted. The source and drain contacts are polarized and  the
direction of the magnetizations $\vec{m}_\alpha$ is indicated by
the arrows.}\label{modelbild}
\end{figure}

Recently, there has been increasing interest in double quantum dot
systems realized e.g. in semiconducting structures \cite{Wiel03}
or carbon nanotubes \cite{Gräber06}, as tunable systems attractive
for studying fundamental spin correlations. In fact the exchange
Coulomb interaction induces a singlet-triplet splitting which can
be used to perform logic gates \cite{Loss98}. Moreover, Coulomb
interaction together with the Pauli principle  can be used to
induce spin-blockade when the two electrons have triplet
correlations
  \cite{Ono02,Johnson05,Liu05,Fransson06}. The Pauli spin-blockade effect can be used
  to obtain a spin-polarized current even in the absence of spin-polarized
  leads; it requires a strong asymmetry between the two on-site energies of
the left and right dot.

  So far transport through a DD system with
  spin-polarized leads has been  addressed  in few theoretical
\cite{Fransson06b,Weymann07,tanaka} and experimental
\cite{Pioro-Ladriere03} works,  for the case of collinearly
polarized leads only. While Ref. \cite{Fransson06b} addresses
additional Pauli spin-blockade regimes when one lead is
half-metallic and one is non-magnetic, Ref. \cite{Weymann07}
focusses on the effects of higher order processes in symmetric DD
systems, which can e.g. yield a zero bias anomaly or a negative
tunneling magnetoresistance. In \cite{Pioro-Ladriere03} Coulomb
blockade spectroscopy is used to measure the energy difference
between symmetric and antisymmetric molecular states, and to
determine the spin of the transferred electron.

In this work we investigate spin-dependent transport in the so-far
unexplored case of a double-dot (DD) system  connected to leads
with arbitrary polarization direction. Specifically, we focus on
the low transparency regime, where a weak coupling between the DD
and the leads is assumed. Our model takes into account interface
reflections as well as exchange effects due to the interactions
and relevant for noncollinear polarization. We focus on the case
of a symmetric DD, so that rectification effects induced by Pauli
spin-blockade  are excluded. In the linear transport regime the
conductance is calculated in closed analytic form. This yields
four distinct resonant tunneling regimes, but due to the
electron-hole symmetry of the DD Hamiltonian, each posses  a
symmetric mirror with respect to the charge neutrality point.
However, by applying an external magnetic field, this symmetry is
broken, which can lead to negative tunneling magnetoresistance
features. Finally, in the nonlinear
 regime some excitation lines can be suppressed for specific
 polarization angles, and negative differential features also
 occur.

 The method developed in this
 work to investigate charge and spin transport
 is based on the Liouville equation for the reduced density matrix
 (RDM) in lowest order in the reflection and tunneling Hamiltonians.
 The obtained equations of motion are fully equivalent to those
 that could be obtained by using the Green's function method \cite{König03,König96}
 in the same weak-tunneling limit. The advantage of our approach
 is that it is,
 in our opinion, easier to understand and to apply  for newcomers,
 as it is based on standard perturbation theory and does
 not require knowledge of the nonequilibrium Green's function
 formalism.

 The paper is organized as follows. In Sec. \ref{model} we
introduce the model system for the ferromagnetic DD
single-electron transistor. In Sec. \ref{dynamics} the coupled
equations of motion for the elements of the DD reduced density
matrix are derived. Readers not interested in the derivation of
the dynamical equations can directly go to Secs. \ref{low bias}
and \ref{high bias}, where results for charge and spin transfer in
the linear and nonlinear regime, respectively, are discussed.
Finally, we present results for the transport characteristics in
the presence of an external magnetic
 field in Sec. \ref{magnetic}. Conclusions are drawn in Sec.
 \ref{conclusions}.

\section{The model}\label{model}
We consider a two-level double-dot (DD), or a single molecule with
two localized atomic orbitals, attached to ferromagnetic source
and drain contacts and with a capacitive coupling to a lateral
gate electrode. The system is described by the total Hamiltonian
\begin{equation}\label{fullhamiltonian}
\hat{H}=\hat{H}_{\odot}+\hat{H_{s}}+\hat{H_{d}}+\hat{H}_{T}+\hat{H}_{R},
\end{equation}
accounting for the DD Hamiltonian, the source $(s)$ and drain
$(d)$ leads, and the tunneling and reflection Hamiltonians,
respectively.
The two contacts are considered to be magnetized along an
arbitrary, but fixed direction determined by the magnetization
vectors $\vec{m}_\alpha$. The two magnetization axes enclose an
angle $\Theta\in[0^\circ,180^\circ]$ (see figure \ref{modelbild}).
 The spin quantization axis $\vec{z}_\alpha$ in lead $\alpha$ is
parallel to the magnetization $\vec{m}_\alpha$ of the lead. The
majority of electrons in each contact will then be in the spin-up
state. The Hamiltonians $\hat{H_{s}},\hat{H_{d}}$ that model the
source ($s$) and drain ($d$) contacts read ($\alpha=s,d$)
\begin{equation}
\hat{H}_\alpha = \sum_{k\sigma_{\alpha}}(\varepsilon_{
k\sigma_\alpha}-\mu_\alpha)c^\dagger_{\alpha k \sigma_\alpha}
c^{\phantom{\dagger}}_{\alpha k \sigma_\alpha},
\end{equation}
where $c^\dagger_{\alpha k \sigma_\alpha}$ and
$c^{\phantom{\dagger}}_{\alpha k \sigma_\alpha}$ are electronic
lead operators. They create, respectively annihilate, electrons
with momentum $k$ and spin $\sigma_{\alpha}$ in lead $\alpha$. The
electrochemical potentials $\mu_\alpha=\mu_{0 \alpha}+eV_\alpha$
contain the bias voltages $V_s$ and $V_d$ at the left and right
lead  with $V_s-V_d=V_{bias}$. There is no voltage drop within the DD. We denote in the following
$\varepsilon_{ k\sigma_\alpha}-\mu_\alpha:=\varepsilon_{\alpha
k\sigma_\alpha}$.

Tunneling processes into and out of the DD  are described by
$\hat{H}_T$. We denote with $d^\dagger_{\alpha \sigma_\alpha}$,
$d^{\phantom{\dagger}}_{\alpha \sigma_\alpha}$ the creation and
destruction operators in the DD. We assume that tunneling only can
happen between a contact and the closest dot, so that we can use
the convention that to the leads indices $\alpha=s,d$  correspond
$\alpha$ = 1,2 for the DD. With $t_\alpha$ the tunneling amplitude
we find
\begin{equation}
\hat{H}_T=\sum_{\alpha k \sigma_\alpha}(t_\alpha
d^\dagger_{\alpha\sigma_\alpha}c^{\phantom{\dagger}}_{\alpha
k\sigma_\alpha}+{t_\alpha^\ast} c^\dagger_{\alpha
k\sigma_\alpha}d^{\phantom{\dagger}}_{\alpha\sigma_\alpha}).
\end{equation}

\begin{figure}
 \centering
 \includegraphics[width=4cm]{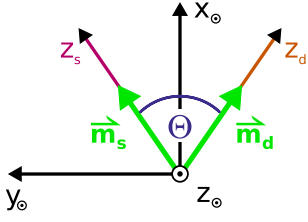}
 \caption{(Color online) The spin quantization axis of the double-dot, $z_\odot$, is chosen to be
  perpendicular to the plane spanned by the magnetization directions $\vec{m}_s$, $\vec{m}_d$ in the leads. The
   latter enclose an angle $\Theta$.}\label{systemkoordinaten}
\end{figure}

The so-called reflection-Hamiltonian $\hat{H}_R$ includes
reflection events at the lead-molecule-interface
\cite{Wetzels05,Koller07}. For strongly shielded leads the overall
effect is the occurrence of a small energy shift $\Delta_R$,
induced  by the magnetic field in the contacts and built up during
several cycles of reflections at the boundaries. It reads
\begin{equation}
\hat{H}_R=-\Delta_R\sum_{\alpha=s,d}(d^\dagger_{{\alpha}\uparrow_\alpha}d^{\phantom{\dagger}}_{{\alpha}\uparrow_\alpha}-d^\dagger_{{\alpha}
\downarrow_\alpha}d^{\phantom{\dagger}}_{{\alpha}\downarrow_\alpha}).
\end{equation}
Finally, the DD Hamiltonian needs to be specified. As spin
quantization axis of the DD, $\vec{z}_\odot$, we  choose  the
direction perpendicular to the plane spanned by $\vec{z}_s$ and
$\vec{z}_d$ \cite{König03}  (see Fig. \ref{systemkoordinaten}).
The two remaining basis vectors $\vec{x}_\odot$ and
$\vec{y}_\odot$ are along $\vec{z}_s+\vec{z}_d$, respectively
$\vec{z}_s-\vec{z}_d$.
The matrices which mathematically describe the above
transformations read

\begin{equation}
M_{s\leftrightarrow\odot}=\frac{1}{\sqrt{2}}
\left(\begin{array}{cc}
    +e^{+i\Theta/4} & +e^{-i\Theta/4} \\
    -e^{+i\Theta/4} & +e^{-i\Theta/4}
\end{array}\right)=M_{d\leftrightarrow\odot}^*.
\end{equation}
%
%

 We express the DD Hamiltonian in the localized basis, such that e.g. $|+,-\rangle$
describes a state with a spin-up electron on site 1 and a spin
down electron on site 2 (with spin directions expressed in the
spin-coordinate system of the DD). Such state can be obtained by
applying  creation operators  on the vacuum state, i.e.,
$|+,-\rangle=d^\dagger_{1\uparrow}d^\dagger_{2\downarrow}|0\rangle$.
In general, the ordering of the creation operators is defined as
$d^\dagger_{1\uparrow}d^\dagger_{2\uparrow}d^\dagger_{1\downarrow}d^\dagger_{2\downarrow}|0\rangle$.
The DD Hamiltonian then reads
\begin{eqnarray}
\hat{H}_{\odot}&=& \sum_{\alpha \sigma_\odot} \varepsilon_\alpha
d^\dagger_{\alpha\sigma_\odot}d^{\phantom{\dagger}}_{\alpha\sigma_\odot}
+
b\sum_{\sigma_\odot}(d^\dagger_{1\sigma_\odot}d^{\phantom{\dagger}}_{2\sigma_\odot}
+d^\dagger_{2\sigma_\odot}d^{\phantom{\dagger}}_{1\sigma_\odot}) \nonumber \\
&+&
(\xi-\frac{U}{2}-V)\sum_{\alpha=1}^2\sum_{\sigma_\odot}d^\dagger_{\alpha\sigma_\odot}
d^{\phantom{\dagger}}_{\alpha\sigma_\odot}
\\
&+&U\sum_{\alpha=1}^2n_{\alpha\uparrow_\odot}n_{\alpha\downarrow_\odot}+V(n_{1\uparrow_\odot}+
n_{1\downarrow_\odot})(n_{2\uparrow_\odot}+n_{2\downarrow_\odot}),\nonumber
\end{eqnarray}
where the spin-index $\odot$ indicates that the operators are
expressed in the spin-coordinate system of the DD. The tunneling
coupling between the two sites is $b$, while $U$ and $V$ are
on-site and inter-site Coulomb interactions. In the remaining we
consider a \emph{symmetric} DD with equal on site energies
$\varepsilon_1=\varepsilon_2$. Thus we can incorporate the on-site
energies in the parameter $\xi$  proportional to the applied gate
voltage $V_{gate}$.

To understand transport properties of the two-site system in the
weak tunneling regime, we have to analyze the eigenstates of the
isolated interacting system. These states, expressed in terms of
the localized states, and the corresponding eigenvalues are listed
in table \ref{table:eigenstatests} \cite{Bulka04}. The table also
indicates the eigenvalues of the total spin operator. The
groundstates of the DD with odd particle number are spin
degenerate.
 In contrast, the groundstates with even particle number have total
 spin $S=0$ and are not degenerate. In the case of the two particles
groundstate the parameters $\alpha_0$ and $\beta_0$ determine
whether the electrons prefer two pair in the same dot or are
delocalized over the DD structure. Since the eigenstates are
normalized to one, the condition $\alpha_0^2+\beta_0^2=1$ holds.
The energy difference between the $S=0$ groundstate and the
triplet is given
 by the exchange energy \[J=\frac{1}{2}(\Delta-U+V)=2|b|(R+\sqrt{1+R^2})\,,\] where
 $\Delta=4|b|\sqrt{1+R^2}$ and $R=(U-V)/(4|b|)$.\\
 Besides the triplet, one observes the presence of higher two-particles
 excited states with total spin $S=0$.

%
%
%
\begin{table}[h!]
\centering
\begin{tabular}{ccccc}\\\\
Abbr.&State&Eigenvalue&Spin\\ \hline\hline \\ 

$|0\rangle$& $|0,0\rangle$ & 0 &$0$  \\ \\ \hline \\

$|1e \sigma\rangle$&$\frac{1}{\sqrt{2}}\left(|\sigma,0\rangle + |0,\sigma\rangle \right)$ & $\xi' + b$  & $1/2$\\
$|1o \sigma\rangle$&$\frac{1}{\sqrt{2}}\left(|\sigma,0\rangle - |0,\sigma\rangle \right)$ & $\xi' - b$  & $1/2$\\ \\
\hline \\ 
$|2\rangle$& $\begin{array}{c}
\frac{\alpha_0}{\sqrt{2}}\left(|+,-\rangle + |-,+\rangle \right)\\
+\frac{\beta_0}{\sqrt{2}}\left(|2,0\rangle+ |0,2\rangle \right)
\end{array} $ &$2\xi'+\frac{1}{2}(U+V-\Delta)$ &$0$\\\\

$|2'(1)\rangle$ &
$|+,+\rangle$& & \\
$|2'(0)\rangle$ & $\frac{1}{\sqrt{2}} \left(|+,-\rangle -
|-,+\rangle \right)$ & $2\xi'+V$ & $1$\\  $|2'(-1)\rangle$ &
$|-,-\rangle$ & &
\\ \\
$|2''\rangle$&
$\frac{1}{\sqrt{2}}\left(|2,0\rangle - |0,2\rangle \right)$&$2\xi'+U$ &$0$\\ \\
$|2'''\rangle$& $\begin{array}{c}
\frac{\beta_0}{\sqrt{2}}\left(|+,-\rangle + |-,+\rangle \right)\\
-\frac{\alpha_0}{\sqrt{2}}\left(|2,0\rangle + |0,2\rangle \right)
\end{array} $ &$2\xi'+\frac{1}{2}(U+V+\Delta)$ &$0$\\\\
\hline \\

$|3o \sigma\rangle$&$\frac{1}{\sqrt{2}}\left(|2,\sigma\rangle+ |\sigma,2\rangle\right)$ & $3\xi'+U+2V+ b$&$1/2$\\
$|3e \sigma \rangle $& $\frac{1}{\sqrt{2}}\left(|2,\sigma\rangle - |\sigma,2\rangle\right)$ & $3\xi'+U+2V- b$&$1/2$\\ \\
\hline \\
$|4\rangle$&
$|2,2\rangle$ &$4\xi'+2U+4V$&$0$\\ \\ \hline\hline\\
\end{tabular}\\
in terms of $R=(U-V)/(4|b|)$:\\\vskip 0.2cm
$\Delta=4|b|\sqrt{1+R^2},\quad
\alpha_0=\frac{1}{\sqrt{2}}\frac{1}{\sqrt{1+R^2-R\sqrt{1+R^2}}}$ \caption{Eigenstates of
the double-dot system and corresponding eigenvalues and parity. In the limit $|b|\to\infty$, where the inter-dot hopping is unhindered, $R\to0$ and $\alpha_0\to\beta_0$. For $|b|\to0$, i.e. no inter-dot hopping takes place, we find, if $U>V$, that $R\to+\infty$ and $\alpha_0\to1$, $\beta_0\to0$; the state $|2\rangle$ then becomes degenerate to $|2'(0)\rangle$, forming a Heitler-London state. In turn, if $U<V$ then $R\to-\infty$ and $\alpha_0\to0$, $\beta_0\to1$.}
\label{table:eigenstatests}
\end{table}

Finally, we remark that $\hat{H}_T$ and $\hat{H}_R$ contain
operators of the DD, $d^\dagger_{\alpha\sigma_\alpha}$ and
$d^{\phantom{\dagger}}_{\alpha\sigma_\alpha}$, with spin
quantization axis of  the leads, while $\hat{H}_{\odot}$ is
already expressed in terms of DD operators
$d^\dagger_{\alpha\sigma_\odot}$ and
$d^{\phantom{\dagger}}_{\alpha\sigma_\odot}$ with spin expressed
in the coordinate system of the DD.\\

%

\section{Dynamical equations for the reduced density
matrix}\label{dynamics}
 In this section we shortly outline how to
derive the equation of motion for the reduced density matrix (RDM)
to lowest non vanishing order in the tunneling and reflection
Hamiltonians. The method is based on the well known Liouville
equation for the total density matrix in lowest order in the
tunneling and reflection Hamiltonian. Equations of motion for the
reduced density matrix are obtained upon performing the trace over
the leads degrees of freedom \cite{Blum96}, yielding,  after
standard approximations, Eqs. (\ref{rhodot}) and (\ref{rhonm})
below. In the case of spin-polarized leads, however, it is
convenient to express the equations of motion for the RDM in the
basis which diagonalizes the isolated system's Hamiltonian {\em
and} in the system's spin quantization axis. After rotation from
the leads´ quantization axis to the DD one, Eq. (\ref{master}),
which forms the basis of all the subsequent analysis, is obtained.

Let us start from the Liouville-equation for the total density
matrix $\hat{\rho}^I(t)$ in the interaction-picture
\begin{equation}\label{liou}
i\hbar\frac{d\hat{\rho}^I(t)}{dt}=[\hat{H}^I_T(t)+\hat{H}^I_R(t),\hat{\rho}^I(t)],
\end{equation}
with $\hat{H}_T$ and $\hat{H}_R$ transformed into the interaction
picture by
$\hat{H}^I_{T/R}(t)=e^{\frac{i}{\hbar}(\hat{H}_{\odot}+\hat{H}_s+\hat{H}_d)(t-t_0)}
\quad \hat{H}_{T/R}\quad
e^{-\frac{i}{\hbar}(\hat{H}_{\odot}+\hat{H}_s+\hat{H}_d)(t-t_0)}$,
where $t_0$ indicates the time at which the perturbation is
switched on.
%
Integrating (\ref{liou}) over time and inserting the obtained
expression in the r.h.s. of (\ref{liou}) one finds equivalently
\begin{eqnarray}\label{rhoint2}
\lefteqn{\hat{\dot{\rho}}^I(t)=-\frac{i}{\hbar}[\hat{H}^I_R(t),\hat{\rho}^I(t_0)]
-\frac{i}{\hbar}[\hat{H}^I_T(t),\hat{\rho}^I(t_0)]}\\
 \nonumber
&-&\frac{1}{\hbar^2} \int^t_{t_0} dt^\prime
[\hat{H}^I_T(t)+\hat{H}^I_R(t),[\hat{H}^I_T(t^\prime)+\hat{H}^I_R(t^\prime),\hat{\rho}^I(t^\prime)]]
 .\end{eqnarray}
The time evolution of the reduced density matrix (RDM)
\begin{equation}\label{rdm}
\hat{\rho}^I_\odot(t):=Tr_{leads}(\hat{\rho}^I(t))
\end{equation} is now
formally obtained from (\ref{rhoint2}) by tracing out the lead
degrees of freedom. To proceed, we  make the following standard
approximations:

i) The leads are considered as reservoirs of non-interacting
electrons which stay in thermal equilibrium at all times. In fact,
we only consider weak tunneling and therefore the influence of the
DD on the leads is marginal. Hence we can factorize the density
matrix of the total system approximatively as
\begin{equation}\label{factorrho}
\hat{\rho}^I(t)\approx\hat{\rho}^I_\odot(t)\hat{\rho}_s\hat{\rho}_d,
\end{equation}
where $\rho_s$ and $\rho_d$ are time independent and given by the
usual thermal equilibrium expression for the contacts $
\hat{\rho}_{s/d}=\frac{e^{-\beta \hat{H}_{s/d}}}{Z_{s,d}}$, with
$\beta$ being the inverse temperature and $Z_{s/d}$ the partition
sums over all states of lead $s/d$.

ii) We consider the lowest non vanishing order in $\hat H_{T/R}$.

iii) We apply the Markov approximation, i.e., in the integral in
Eq. (\ref{rhoint2}) we replace $\hat{\rho}^I_\odot(t^\prime)$ with
$\hat{{\rho}}^I_\odot(t)$.
 In other words, it is assumed that the system looses all memory of its
past due to the interaction with the leads electrons.

 Furthermore, being interested
in the long term behavior of the system only, we send
$t_0\rightarrow -\infty$.
We finally obtain the generalized
master equation (GME) for the reduced density matrix
\begin{eqnarray}\label{GME}
\lefteqn{\hat{\dot{\rho}}^I_\odot(t)=-\frac{i}{\hbar}Tr_{leads}[\hat{H}^I_R(t),\hat{\rho}^I_\odot(t)\hat{\rho}_s\hat{\rho}_d]}\\
&-&\frac{1}{\hbar^2}\int^\infty_0 dt^{\prime\prime}
Tr_{leads}\left(
[\hat{H}^I_T(t),[\hat{H}^I_T(t-t^{\prime\prime}),\hat{\rho}^I_\odot(t)\hat{\rho}_s\hat{\rho}_d]]\right).
\nonumber
\end{eqnarray}
%
%
\subsection{Contribution from the tunneling Hamiltonian }
In the following, we derive the explicit expression for the GME in
the basis of the isolated DD.
 For
simplicity we omit the contribution of the reflection Hamiltonian
in a first instance. When we shall have obtained the final form of
the GME due to the tunneling term, we will see that it is easy to
insert the contribution from  the reflection Hamiltonian.
Let us then start from the tunneling Hamiltonian in the
interaction picture
\begin{eqnarray}\label{tunnelinter}
\lefteqn{\hat{H}_T^I(t)=\sum_{\alpha k \sigma_\alpha}\sum_{i,j}}
\\
\nonumber && t_\alpha c^\dagger_{\alpha k \sigma_\alpha}
(d^{\phantom{\dagger}}_{\alpha\sigma_\alpha})_{ij}\ket{i}\bra{j}\exp\left[
i(\varepsilon_i-\varepsilon_j +\varepsilon_{\alpha
k\sigma_\alpha})t/\hbar \right]  + h.c.
\end{eqnarray}
where
$\big(d^{\phantom{\dagger}}_{\alpha\sigma_\alpha}\big)_{ij}=\langle
i\vert d^{\phantom{\dagger}}_{\alpha\sigma_\alpha}\vert j \rangle$
and $\big(d^\dagger_{\alpha\sigma_\alpha}\big)_{ij}=\langle i\vert
d^\dagger_{\alpha\sigma_\alpha}\vert j \rangle$ are the electron
annihilation- and creation-operators in the spin-quantization axis
of lead $\alpha$ expressed in the basis of the energy eigenstates
of the quantum dot system.
To simplify (\ref{GME}) standards approximations are invoked.
 i) The first one is the secular
approximation: Fast oscillations in time
average out in the
stationary limit we are interested in, and thus can be neglected.
 Together with the relation
$Tr_{leads}( \hat{\rho}_s\hat{\rho}_dc^\dagger_{\alpha k
\sigma_\alpha}c_{\alpha^\prime k^\prime\sigma_\alpha^\prime})
=\delta_{kk^\prime}\delta_{\alpha\alpha^\prime}\delta_{\sigma\sigma^\prime}f_\alpha(\varepsilon_{\alpha
k\sigma}), $
where $f_\alpha(\varepsilon_{\alpha k\sigma})$ is the Fermi
function, and the cyclic properties of the trace we get
\begin{widetext}
\begin{eqnarray}
\label{rhodot}\dot{\hat{\rho}}^I_\odot(t)&=&-\frac{1}{\hbar^2}\int^\infty_0
dt^{\prime\prime}
\sum_{\alpha k \sigma_\alpha}\vert{t_{\alpha}}\vert^{2}\left\lbrace\right.  \\
\nonumber&&\hspace{-1.0cm}+\sum_{ilm}f_\alpha(\varepsilon_{\alpha
k\sigma_\alpha})(d^{\phantom{\dagger}}_{\alpha\sigma_\alpha})_{il}
(d^\dagger_{\alpha\sigma_\alpha})_{lm}\ket{i}\bra{m}\hat{\rho}^I_\odot(t)\exp\left[
i(\varepsilon_m-\varepsilon_l+\varepsilon_{\alpha k\sigma_\alpha})t^{\prime\prime}/\hbar\right]
%
%
\\
\nonumber&&\hspace{-1cm}+
\sum_{ilm}(1-f_\alpha(\varepsilon_{\alpha
k\sigma_\alpha}))(d^\dagger_{\alpha\sigma_\alpha})_{il}
(d^{\phantom{\dagger}}_{\alpha\sigma_\alpha})_{lm}\ket{i}\bra{m}\hat{\rho}^I_\odot(t)\exp\left[
-
i(\varepsilon_l-\varepsilon_m+\varepsilon_{\alpha k\sigma_\alpha})t^{\prime\prime}/\hbar\right]\\
\nonumber&&\hspace{-1cm}+\sum_{ilm}f_\alpha(\varepsilon_{\alpha
k\sigma_\alpha})\hat{\rho}^I_\odot(t)
(d^{\phantom{\dagger}}_{\alpha\sigma_\alpha})_{il}(d^\dagger_{\alpha\sigma_\alpha})_{lm}\ket{i}\bra{m}\exp\left[-
i(\varepsilon_i-\varepsilon_l+\varepsilon_{\alpha k\sigma_\alpha})t^{\prime\prime}/\hbar\right]\\
\nonumber&&\hspace{-1cm}+
\sum_{ilm}(1-f_\alpha(\varepsilon_{\alpha k
\sigma_\alpha}))\hat{\rho}^I_\odot(t)
(d^\dagger_{\alpha\sigma_\alpha})_{il}(d^{\phantom{\dagger}}_{\alpha\sigma_\alpha})_{lm}\ket{i}\bra{m}\exp\left[
+
i(\varepsilon_l-\varepsilon_i+\varepsilon_{\alpha k \sigma_\alpha})t^{\prime\prime}/\hbar\right]\\
\nonumber&&\hspace{-1cm}-
\sum_{iljm}(1-f_\alpha(\varepsilon_{\alpha
k\sigma_\alpha}))(d^{\phantom{\dagger}}_{\alpha\sigma_\alpha})_{ij}
\hat{\rho}^I_\odot(t)_{jl}(d^\dagger_{\alpha\sigma_\alpha})_{lm}\ket{i}\bra{m}\exp\left[
+
i(\varepsilon_m-\varepsilon_l+\varepsilon_{\alpha k \sigma_\alpha})t^{\prime\prime}/\hbar\right]\\
\nonumber&&\hspace{-1cm}- \sum_{iljm}f_\alpha(\varepsilon_{\alpha
k\sigma_\alpha})(d^\dagger_{\alpha\sigma_\alpha})_{ij}
\hat{\rho}^I_\odot(t)_{jl}(d^{\phantom{\dagger}}_{\alpha\sigma_\alpha})_{lm}\ket{i}\bra{m}\exp\left[
-
i(\varepsilon_l-\varepsilon_m+\varepsilon_{\alpha k \sigma_\alpha})t^{\prime\prime}/\hbar\right]\\
\nonumber&&\hspace{-1cm}-
\sum_{iljm}(1-f_\alpha(\varepsilon_{\alpha
k\sigma_\alpha}))(d^{\phantom{\dagger}}_{\alpha\sigma_\alpha})_{ij}
\hat{\rho}^I_\odot(t)_{jl}(d^\dagger_{\alpha\sigma_\alpha})_{lm}\ket{i}\bra{m}\exp\left[
-
i(\varepsilon_i-\varepsilon_j+\varepsilon_{\alpha k\sigma_\alpha})t^{\prime\prime}/\hbar\right]\\
\nonumber&&\hspace{-1cm}- \sum_{iljm}f_\alpha(\varepsilon_{\alpha
k
\sigma_\alpha})(d^\dagger_{\alpha\sigma_\alpha})_{ij}\hat{\rho}^I_\odot(t)_{jl}(d^{\phantom{\dagger}}_{\alpha\sigma_\alpha})_{lm}\ket{i}\bra{m}\exp\left[
+
i(\varepsilon_j-\varepsilon_i+\varepsilon_{\alpha k
\sigma_\alpha})t^{\prime\prime}/\hbar\right]\left. \right\rbrace.
\end{eqnarray}
\end{widetext}
ii) For the  second approximation we notice that we wish to
evaluate single components $\bra{n}\hat{\rho}_\odot^I\ket{m}$ of
the RDM in the system's energy eigenbasis. Therefore,  we assume
that the DD is in a pure charge state with a certain number of
electrons $N$ and energy $E_N$ . In fact, in the weak tunneling
limit the time between two tunneling events is longer than the
time where relaxation processes happen. That is, we can neglect
matrix elements between states with different number of electrons,
and only regard elements of $\hat{\rho}_\odot^I$ which connect
states with same electron number $N$ and same energy $E_N$. So we
can divide $\hat{\rho}_\odot^I$ into sub-matrices labelled with
$N$ and $E_N$ and find
\begin{widetext}
\begin{eqnarray}\label{rhonm}
&&\dot{{\rho}}^{E_NN}_{{nm}}(t)=-\frac{\pi}{\hbar}\sum_{\alpha\sigma_\alpha}\quad\left\lbrace
\sum_{l,l^\prime\in{\ket{N-1}}},\quad\sum_{j\in
\ket{E_NN}},\quad\sum_{h,h^{\prime}\in{\ket{N+1}}}\right\rbrace
\vert{t_{\alpha}}\vert^{2} \left\lbrace  \right. \\
\nonumber\mathrm{(a)}&&+\left[ f_\alpha(\varepsilon_h-\varepsilon_j)
D_{\alpha\sigma_\alpha}(\varepsilon_h-\varepsilon_j)
+\frac{i}{\pi}\int^\prime d\varepsilon_{k}\frac{f_\alpha(\varepsilon_{k})
D_{\alpha\sigma_\alpha}(\varepsilon_{k})}{\varepsilon_{k}-\varepsilon_h+\varepsilon_j}\right]
(d^{\phantom{\dagger}}_{\alpha\sigma_\alpha})_{nh}(d^\dagger_{\alpha\sigma_\alpha})_{hj}{\rho}^{E_NN}_{{jm}}(t)\\
\nonumber\mathrm{(b)}&&+\left[ \left( 1-f_\alpha(\varepsilon_j-\varepsilon_l)\right)
D_{\alpha\sigma_\alpha}(\varepsilon_j-\varepsilon_l)-\frac{i}{\pi}\int^\prime d\varepsilon_{k}
\frac{\left( 1-f_\alpha(\varepsilon_{k})\right) D_{\alpha\sigma_\alpha}
(\varepsilon_{k})}{\varepsilon_{k}-\varepsilon_j+\varepsilon_l}\right]
(d^\dagger_{\alpha\sigma_\alpha})_{nl}(d^{\phantom{\dagger}}_{\alpha\sigma_\alpha})_{lj}{\rho}^{E_NN}_{{jm}}(t)\\
\nonumber\mathrm{(c)}&&+\left[ f_\alpha(\varepsilon_h-\varepsilon_j)D_{\alpha\sigma_\alpha}
(\varepsilon_h-\varepsilon_j)-\frac{i}{\pi}\int^\prime d\varepsilon_{k}\frac{f_\alpha(\varepsilon_{k})
D_{\alpha\sigma_\alpha}(\varepsilon_{k})}{\varepsilon_{k}-\varepsilon_h+\varepsilon_j}\right] {\rho}^{E_NN}_{{nj}}(t)(d_{\alpha\sigma_\alpha})_{jh}(d^\dagger_{\alpha\sigma_\alpha})_{hm}\\
\nonumber\mathrm{(d)}&&+\left[ \left( 1-f_\alpha(\varepsilon_j-\varepsilon_l)\right) D_{\alpha\sigma_\alpha}
(\varepsilon_j-\varepsilon_l)+\frac{i}{\pi}\int^\prime d\varepsilon_{k}\frac{\left( 1-f_\alpha(\varepsilon_{k})\right)
D_{\alpha\sigma_\alpha}(\varepsilon_{k})}{\varepsilon_{k}-\varepsilon_j+\varepsilon_l}\right] {\rho}^{E_NN}_{{nj}}(t)
(d^\dagger_{\alpha\sigma_\alpha})_{jl}(d^{\phantom{\dagger}}_{\alpha\sigma_\alpha})_{lm}\\
\nonumber\mathrm{(e)}&&-2 \left(
1-f_\alpha(\varepsilon_h-\varepsilon_j)\right)
D_{\alpha\sigma_\alpha} (\varepsilon_h-\varepsilon_j)
(d^{\phantom{\dagger}}_{\alpha\sigma_\alpha})_{nh^\prime}(d^\dagger_{\alpha\sigma_\alpha})_{hm}
{\rho}^{E_{h}N+1}_{{h^\prime h}}(t)\\
\nonumber\mathrm{(f)}&&-2
f_\alpha(\varepsilon_j-\varepsilon_l)
D_{\alpha\sigma_\alpha}(\varepsilon_j-\varepsilon_l)
(d^\dagger_{\alpha\sigma_\alpha})_{nl^\prime}(d^{\phantom{\dagger}}_{\alpha\sigma_\alpha})_{lm}{\rho}^{E_{l}N-1}_{{l^\prime
l}}(t)
\left. \right\rbrace . 
\end{eqnarray}
\end{widetext}
%
In (\ref{rhonm}) we used the notation ${\rho}^{E_{N}N}_{nm}:=\langle n |{\hat\rho}^{I,E_{N}N}_\odot |m\rangle$.\smallskip\\
By convention, $\lbrace \sum_{l,l^\prime},\sum_{j},
\sum_{h,h^\prime}\rbrace $ means that in each line (a)-(f)
we sum over the indices occurring in this line only. Notice that the sum over $j$ is restricted to states of energy $E_j=E_N=E_n=E_m$. For the states with $N\pm1$ electrons, we have to sum over all energies, therefore we indexed the density matrix with $E_h=E_{h'}$ respectively $E_l=E_{l'}$ in lines (e) and (f).
Further, we replaced the sum over $k$ by an integral: $ \sum_k\longrightarrow
\int d\varepsilon_{\alpha k\sigma_\alpha}
D_{\alpha\sigma_\alpha}(\varepsilon_{\alpha k\sigma_\alpha})$,
where $D_{\alpha\sigma_\alpha}(\varepsilon_{\alpha
k\sigma_\alpha})$ denotes the density of states in lead $\alpha$
for the spin direction $\sigma_\alpha$, and applied  the useful
formula
\begin{eqnarray}
&&\int d\varepsilon_{\alpha k\sigma_\alpha} G(\varepsilon_{\alpha
k\sigma_\alpha}) \int_0^t dt^{\prime\prime}
e^{\pm\frac{i}{\hbar}(\varepsilon_{\alpha
k\sigma_\alpha}-E)t^{\prime\prime}}\nonumber
\\ && = \pi\hbar G(E) \pm i\hbar \int^\prime
d\varepsilon_{\alpha k\sigma_\alpha}\frac{G(\varepsilon_{\alpha
k\sigma_\alpha})}{(\varepsilon_{\alpha k\sigma_\alpha}-E)}\;,
\end{eqnarray}
where the prime at the integral denotes Cauchy's principal part
integration. In our case $G(\varepsilon_{\alpha
k\sigma_\alpha})=D_{\alpha\sigma_\alpha}(\varepsilon_{\alpha
k\sigma_\alpha}) f^\pm_\alpha(\varepsilon_{\alpha
k\sigma_\alpha})$ with $f^+_\alpha=f_\alpha$ and
$f^-_\alpha=1-f_\alpha$. In order to simplify the notations we
replaced $\varepsilon_{\alpha k\sigma_\alpha}$ by $\varepsilon_k$
in (\ref{rhonm}).
\subsection{Transformation into the spin coordinate system of the double-dot}

In the previous section we introduced the transformation rules for
changing from the lead spin coordinates $\sigma_\alpha$ into the
DD spin coordinates $\sigma_\odot$. These rules give
\begin{equation}
\left( \begin{array}{c}
         d^\dagger_{\alpha\uparrow_\alpha}\\
     d^\dagger_{\alpha\downarrow_\alpha}\end{array}\right) =\frac{1}{\sqrt{2}}
\left(\begin{array}{cc}
    +e^{-i\Theta_\alpha/2} & +e^{+i\Theta_\alpha/2} \\
    -e^{-i\Theta_\alpha/2} & +e^{+i\Theta_\alpha/2}
\end{array}\right)\left(\begin{array}{c}
 d^\dagger_{\alpha\uparrow_\odot}\\
     d^\dagger_{\alpha\downarrow_\odot}
\end{array} \right)
\end{equation}
with $\Theta_{s}:=-\frac{\Theta}{2}$, $\Theta_{d}:=+\frac{\Theta}{2}$.\smallskip\\
Thus, Eq. (\ref{rhonm}) can be easily expressed in the DD spin
quantization axis.  For example it holds
\begin{eqnarray}
&&\sum_{\sigma_\alpha}D_{\alpha\sigma_\alpha}d^\dagger_{\sigma_\alpha}d_{\sigma_\alpha}=\\
&&\nonumber\frac{1}{2}\left(D_{\alpha \uparrow_\alpha}+D_{\alpha
\downarrow_\alpha} \right) \sum_{\sigma_\odot}
\Phi_{\alpha\sigma_\odot\sigma_\odot}
d^\dagger_{\alpha\sigma_\odot}d^{\phantom{\dagger}}_{\alpha\sigma_\odot} \nonumber \\
&+&\frac{1}{2}\left(D_{\alpha \uparrow_\alpha}-D_{\alpha
\downarrow_\alpha} \right)\sum_{\sigma_\odot}
\Phi^*_{\alpha\sigma_\odot-\sigma_\odot}
d^\dagger_{\alpha\sigma_\odot}d^{\phantom{\dagger}}_{\alpha-\sigma_\odot}
\nonumber\end{eqnarray}
%
%
where we introduced
\begin{equation*}
\Phi_{\alpha\sigma_\odot\sigma_\odot^\prime}:=\left\{\begin{array}{l@{\qquad
}r}
                    1 & \sigma_\odot=\sigma_\odot^\prime \;,\\
                    e^{i\Theta_\alpha} & \sigma_\odot=\uparrow \quad \sigma_\odot^\prime=\downarrow \;, \\
                    e^{-i\Theta_\alpha} &  \sigma_\odot=\downarrow \quad \sigma_\odot^\prime=\uparrow \;.\\
                             \end{array}\right.
\end{equation*}
For later convenience we also define
\begin{equation*}F^{\pm}_{\alpha\sigma_\odot\sigma_\odot^{\prime}}:=\frac{1}{2}\left\{\begin{array}{l@{\qquad }r}
                        D_{\alpha +_\alpha}f^\pm_\alpha(E)+D_{\alpha -_\alpha}f^\pm_\alpha(E) & \sigma_\odot=\sigma_\odot^\prime,\\
                        D_{\alpha +_\alpha}f^\pm_\alpha(E)-D_{\alpha -_\alpha}f^\pm_\alpha(E) &
                        \sigma_\odot\neq\sigma_\odot^\prime,
                                                                           \end{array}\right.\end{equation*}
and its related principal part integral
\begin{equation*}
P^{\pm}_{\alpha\sigma_\odot\sigma_\odot^{\prime}}(E):=\int^\prime
d{\varepsilon}
F^{\pm}_{\alpha\sigma_\odot\sigma_\odot^{\prime}}(\varepsilon)(\varepsilon-E)^{-1}.
\end{equation*}
\subsection{Contribution from the reflection Hamiltonian}
In order to give the full expression for the GME in the system's
eigenbasis, we need to compute the contribution from the
reflection Hamiltonian in Eq. (\ref{GME}). In analogy to what we
did to evaluate the contribution from the tunneling Hamiltonian,
we must first transform $\hat{H}_R$ into the interaction picture
and then perform the secular approximation to get rid of the
time-dependence. To start, we express $\hat{H}_R$ in the DD spin
quantization basis,
\begin{eqnarray}
\hat{H}^I_R&=&-\Delta_R\sum_\alpha\sum_{\substack{j\in\ket{N}\\l\in\ket{N-1}}}\sum_{\sigma_\odot\neq\sigma_\odot^\prime}
\Phi^\ast_{\alpha\sigma_\odot\sigma_\odot^\prime}{d^\dagger_{\sigma_\odot}}_{jl}{d_{\sigma_\odot^\prime}}_{lj}
\ket{j}\bra{j}.\nonumber\\
\end{eqnarray}
%
The commutator is easily evaluated to be
\begin{eqnarray}
\label{href}
&&-\frac{i}{\hbar}Tr_{leads}[\hat{H}^I_R,\hat{\rho}^I_\odot(t)\rho_s\rho_d]=\\
\nonumber&&-\frac{i}{\hbar}
\sum_\alpha\Delta_R\sum_{j\in\ket{N}}\sum_{l\in\ket{N-1}}
\sum_{\sigma_\odot\neq\sigma_\odot^\prime}\Phi^\ast_{\alpha\sigma_\odot\sigma_\odot^\prime}
\nonumber\\
&& \left[
{d^\dagger_{\sigma_\odot}}_{jl}{d_{\sigma_\odot^\prime}}_{lj}
\ket{j}\bra{j}\hat{\rho}^I_\odot(t)
-\hat{\rho}^I_\odot(t){d^\dagger_{\sigma_\odot}}_{jl}{d_{\sigma_\odot^\prime}}_{lj}
\ket{j}\bra{j}\right] .\nonumber
\end{eqnarray}
 In order to include this
commutator in the master equation (\ref{rhonm}) let us introduce
the  abbreviation
\begin{equation}
R_{\alpha\sigma_\odot\sigma_\odot^\prime}=\frac{1}{\vert{t^{\alpha}}\vert^{2}}\Delta_R\left(\delta_{\sigma_\odot\uparrow}\delta_{\sigma_\odot^\prime\downarrow}+\delta_{\sigma_\odot\downarrow}\delta_{\sigma_\odot^\prime\uparrow}
\right).
\end{equation}
 Now we can add
$R_{\alpha\sigma_\odot\sigma_\odot^\prime}$ in (\ref{rhonm}) in
the lines ($b$) and ($d$) to find the final form of the complete
master equation in the DD spin-coordinate system. It reads
\begin{widetext}
\begin{eqnarray}\label{master}
&& \dot{{\rho}}_{ nm}^{E_{N}N}(t) =
-\frac{\pi}{\hbar}\sum_{\alpha=s,d}\vert{t^{\alpha}}\vert^{2}\sum_{\sigma_{\odot},\sigma\odot^{\prime}}
\left\lbrace\sum_{l,l^\prime\in{\ket{N-1}}},\quad\sum_{j\in\ket{E_NN}},\quad\sum_{h,h^{\prime}\in{\ket{N+1}}}\right\rbrace
\left\lbrace \right.
\\\nonumber
\mathrm{(a)}&&+\Phi_{\alpha\sigma_\odot\sigma_\odot^\prime}\big[F^{+}_{\alpha\sigma_\odot\sigma_\odot^{\prime}}
(\varepsilon_{h}-\varepsilon_{j})+\frac{i}{\pi}P^{+}_{\alpha\sigma_\odot\sigma_\odot^{\prime}}
(\varepsilon_{h}-\varepsilon_{j})\big]\big(d_{\alpha\sigma_\odot}\big)_{nh}
\big(d^{\dagger}_{\alpha\sigma_\odot^\prime}\big)_{hj}\;{\rho}_{
jm}^{E_{N} N}(t)\end{eqnarray}
\begin{eqnarray*}
\mathrm{(b)}&&+\Phi^\ast_{\alpha\sigma_\odot\sigma_\odot^\prime}\Big[F^{-}_{\alpha\sigma_\odot\sigma_\odot^{\prime}}
(\varepsilon_{j}-\varepsilon_{l})-\frac{i}{\pi}\left[
P^{-}_{\alpha\sigma_\odot\sigma_\odot^{\prime}}(\varepsilon_{j}-\varepsilon_{l})
+R_{\alpha\sigma_\odot\sigma^{\prime}_\odot}\right]
\Big]\big(d^{\dagger}_{\alpha\sigma_\odot}\big)_{nl}\big(d_{\alpha\sigma_\odot^\prime}\big)_{lj}{\rho}_{
jm}^{E_{N} N}(t)\\\nonumber
\mathrm{(c)}&&+\Phi_{\alpha\sigma_\odot\sigma_\odot^\prime}
\big[F^{+}_{\alpha\sigma_\odot\sigma_\odot^{\prime}}(\varepsilon_{h}-\varepsilon_{j})-\frac{i}{\pi}
P^{+}_{\alpha\sigma_\odot\sigma_\odot^{\prime}}(\varepsilon_{h}-\varepsilon_{j})\big]{\rho}
_{
nj}^{E_{N} N}(t)\big(d_{\alpha\sigma_\odot}\big)_{jh}\big(d^{\dagger}_{\alpha\sigma_\odot^\prime}\big)_{hm}\\
\nonumber
\mathrm{(d)}&&+\Phi^\ast_{\alpha\sigma_\odot\sigma_\odot^\prime}
\Big[F^{-}_{\alpha\sigma_\odot\sigma_\odot^{\prime}}
(\varepsilon_{j}-\varepsilon_{l})+\frac{i}{\pi}\left[
P^{-}_{\alpha\sigma_\odot\sigma_\odot^{\prime}}(\varepsilon_{j}-\varepsilon_{l})
+R_{\alpha\sigma_\odot\sigma^{\prime}_\odot}\right] \Big]{\rho}_{
nj}^{E_{N}
N}(t)\big(d^{\dagger}_{\alpha\sigma_\odot}\big)_{jl}\big(d_{\alpha\sigma_\odot^\prime}\big)_{lm}\\\nonumber
\mathrm{(e)}&&-2\Phi_{\alpha\sigma_\odot\sigma_\odot^\prime}F^{-}_{\alpha\sigma_\odot\sigma_\odot^{\prime}}
(\varepsilon_{h}-\varepsilon_{j})\big(d_{\alpha\sigma_\odot}\big)_{nh^\prime}
\rho_{ h^{\prime}h}^{E_{h}
N+1}(t)\big(d^{\dagger}_{\alpha\sigma_\odot^\prime}\big)_{hm}\\\nonumber
\mathrm{(f)}&&-2\Phi^\ast_{\alpha\sigma_\odot\sigma_\odot^\prime}
F^{+}_{\alpha\sigma_\odot\sigma_\odot^{\prime}}(\varepsilon_{j}-\varepsilon_{l})
\big(d^\dagger_{\alpha\sigma_\odot}\big)_{nl^\prime}\rho_{
l^{\prime}l}^{E_{l} N-1}(t)
\big(d_{\alpha\sigma_\odot^\prime}\big)_{lm}\left. \right\rbrace .
\end{eqnarray*}
\end{widetext}
%
\subsection{The current formula}
We observe now that (\ref{master}) can be recast in   the
Bloch-Redfield form
\begin{eqnarray}
\dot{\rho}_{ nm}^{E_N N}(t) &=& - \sum_{jj'}R_{nm\, jj'}^{N N}\
\rho_{jj'}^{E_N N}(t)\nonumber \\
&&+ \sum_{hh'} R_{nm\, hh'}^{N\, N+1}\rho_{ hh'}^{E_{h}
N+1}(t)\nonumber \\
&&+ \sum_{ll'} R_{nm\, ll'}^{N\, N-1}\rho_{ ll'}^{E_{l} N-1}(t)
,\label{eq:generalized_Mequ}\end{eqnarray} where the sums in
(\ref{eq:generalized_Mequ}) run over states with
fixed particle number: $j,j'\in \{\left|E_N N\right. \rangle\}$, $h,h'\in
\{\left| N+1 \right. \rangle\}$, $l,l'\in \{\left| N-1 \right.
\rangle\}$. The Redfield tensors are given by $(\alpha =s,d)$
\cite{Koller07}
 \begin{eqnarray}
R_{nm\, jj'}^{N N} & = & \sum_{\alpha}\sum_{l (\texttt{or}\; h)}
\left[\delta_{mj'} \left(\Gamma_{\alpha, nhhj}^{(+)N\, N+1}+
\Gamma_{\alpha, nllj}^{(+)N\, N-1}\right)\right. \nonumber\\
&&+ \left. \delta_{nj}\left(\Gamma_{\alpha , j'hhm}^{(-)N
N+1}+\Gamma_{\alpha , j'llm}^{(-)N N-1}\right)\right]
\label{eq:Blochredf_1},
\end{eqnarray}
\begin{equation}
 R_{nm\, kk'}^{N\,N\pm 1}=\sum_{\alpha} \left(\Gamma_{\alpha ,
k'mnk}^{(+) N N\mp 1} + \Gamma_{\alpha , k'mnk}^{(-) N N\mp 1}
\right), \end{equation}
 where  the
quantities $\Gamma_{\alpha, njjk}^{(\pm)N\,N\pm 1}$  can be easily
read out from (\ref{master}). They are
\begin{eqnarray}\Gamma_{\alpha , nhh'k}^{(\pm)N N+1}&=&\sum_{\sigma_\odot
\sigma'_\odot}\{\frac{\pi}{\hbar}\Phi_{\alpha \sigma_\odot
\sigma'_\odot}|t_\alpha|^2[F^+_{\alpha \sigma_\odot \sigma'_\odot}
(\varepsilon_h -
\varepsilon_k) \nonumber \\
 &\pm& \frac{i}{\pi}P^+_{\alpha \sigma
\sigma' }(\varepsilon_h - \varepsilon_k) ](d_{\alpha
\sigma_\odot})_{nh}(d^\dagger_{\alpha \sigma'_\odot})_{h'k}\},
\nonumber
\end{eqnarray}
\begin{eqnarray}
\lefteqn{\Gamma_{\alpha , nll'k}^{(\pm)N N-1} = \sum_{\sigma_\odot
\sigma'_\odot}\{\frac{\pi}{\hbar}\Phi^*_{\alpha \sigma_\odot
\sigma'_\odot}|t_\alpha|^2[F^-_{\alpha \sigma_\odot \sigma'_\odot}
(\varepsilon_k - \varepsilon_l)} \nonumber \\&\mp&
\frac{i}{\pi}(P^-_{\alpha \sigma_\odot \sigma'_\odot
}(\varepsilon_k - \varepsilon_l)+R_{\alpha \sigma_\odot
\sigma'_\odot}) ](d^\dagger_{\alpha \sigma_\odot})_{nl}(d_{\alpha
\sigma'_\odot})_{l'k}\}\;.\nonumber
\end{eqnarray}
 With the stationary density matrix $\hat\rho_{\odot
st}^{I}$ being known, the current (through lead $\alpha=s/d=\pm$)
follows from
\begin{equation}
I= 2\alpha e {\rm Re}\sum_{N}\sum_{n,n',j} \left(\Gamma_{\alpha,
njjn'}^{(+)N\, {N+1}}-\Gamma_{\alpha, njjn'}^{(+)N\,
{N-1}}\right)\rho_{n'n,st}^{E_n N}.\label{current}
\end{equation}
We solve Eq. (\ref{eq:generalized_Mequ}) numerically and use the
result to evaluate the current flowing through the DD, as shown in
the forthcoming sections.
 At low bias voltages,  however, we can make some further approximations to arrive at an
analytical formula for the static DC current.

\section{The low-bias regime}\label{low bias}
\subsection{General considerations}

A low bias voltage ensures that merely one channel is involved
with respect to transport properties. Here we focus on gate
voltages which align charge states $N$ and $N+1$. Moreover, we can
focus on density matrix elements  which involve the energy ground
states $E_N^{(0)}$ and $E_{N+1}^{(0)}$ only.
In the following we shall use the compact notations
\begin{equation}
\hat\rho_\odot^{I,E_N^{(0)}N}:=\hat\rho_\odot^{(N)};\quad \langle
n|\hat\rho_\odot^{(N)}|m\rangle =\rho_{nm}^{(N)} \;.
\end{equation}

Evaluation of the current requires the knowledge of
$\hat{\rho}_\odot^{(N)}$ and $\hat{\rho}_\odot^{(N+1)}$, i.e. a
solution of the set of coupled equations obtained from (\ref{master}), or,
equivalently, from (\ref{eq:generalized_Mequ}). In the low bias
regime this task is simplified since i) terms which try to couple
states with particle numbers unlike $N$ and $N+1$ can be
neglected; ii) we can reduce the sums over $h, h^\prime$ in the
equation for $\dot{\hat{\rho}}_\odot^{(N)}$, and over $l,
l^\prime$ in the equation for $\dot{\hat{\rho}}_\odot^{(N+1)}$ to
energy-groundstates $E_N^{(0)}$ and $E_{N+1}^{(0)}$, because all
the other transitions are suppressed exponentially by the
Fermi-function. Notice, however, that these two approximations are
not appropriate for the principal-part-terms, since they are not
energy conserving. The resulting equations for
$\hat{\rho}_\odot^{(N)}$ and $\hat{\rho}_\odot^{(N+1)}$, Eqs.
(\ref{Gl N}) and (\ref{Gl N+1}) respectively, can be found in
appendix \ref{appendix1}.
%
%
%
In the following we shall apply those equations to derive an
analytical expression for the conductance in the four different
resonant charge state regimes possible in a DD system, i.e.,
 \begin{eqnarray} N&=&0 \leftrightarrow N=1, \qquad
        N = 1 \leftrightarrow N=2,\nonumber  \\
        N&=&2 \leftrightarrow N = 3, \qquad  N=3 \leftrightarrow N = 4.\label{resonances}
           \end{eqnarray}
In all of the four cases we get a system of five coupled equations
involving diagonal and off-diagonal elements of the RDM. The
matrix elements of the dot operators between the involved states
entering these equations are given in appendix  \ref{appendix2}.
Before going into the details of these equations, it is
instructive to analyze the structure and the physical significance
of the involved RDM elements.

\subsection{The elements of the reduced density matrix }
%
%
 \textbf{N = 0}. In the case of an empty system we have only one density matrix
element in the corresponding block with fixed particle number
$N=0$, i.e.,
\begin{equation}
{\rho}^{(0)}_{00}(t) =:W_0,
\end{equation}
 describing the probability to find an empty double-dot system.\\

\textbf{N = 1}. In this case we have four eigenstates for the
system, where the two {\em even} ones build the degenerate
groundstate and the two {\em odd} ones are excited states (see
table \ref{table:eigenstatests}). In the low-bias-regime we only
need to take into account transitions between groundstates.
Therefore we have to deal with the 2 by 2 matrix
\begin{equation}
\left(\begin{array}{cc}
    {\rho}^{(1)}_{ 1e\uparrow 1e\uparrow} & {\rho}^{(1)}_{ 1e\uparrow 1e\downarrow}\\
     {\rho}^{(1)}_{ 1e\downarrow 1e\uparrow} & {\rho}^{(1)}_{ 1e\downarrow 1e\downarrow}
\end{array}\right)
=:\left(\begin{array}{cc}
     W_{1\uparrow} & w_1e^{i \alpha_1}\\
     w_1e^{-i\alpha_1} & W_{1\downarrow}
\end{array}\right).
\end{equation}
The total occupation probability for one electron is
\begin{equation}
W_1:=W_{1\uparrow}+W_{1\downarrow}.
\end{equation}
The meaning of the off-diagonal elements, the so called
coherences, becomes clear if we regard the average spin in the
system
\begin{equation}
S^{(1)}_i=\frac{1}{2}Tr\left(\sigma^{Pauli}_i\hat{\rho}^{(1)}_\odot(t)\right)\;,
\end{equation}
where $i={x,y,z}$ and $\sigma^{Pauli}_i$ are the Pauli
spin-matrices. This yields
\begin{eqnarray}\label{S1}
S^{(1)}_x&=& w_1 \cos{\alpha_1},\quad
S^{(1)}_y= -w_1 \sin{\alpha_1},\\
S^{(1)}_z&=&\frac{1}{2}(W_{1\uparrow}-W_{1\downarrow}).
\end{eqnarray}
\textbf{N = 2}.  For the case $N=2$ we actually have six different
eigenstates, but only one of them, $\ket{2}$, is a groundstate
(with spin $S=0$), see table \ref{table:eigenstatests}. Only this
groundstate must be considered in the low-bias-regime, yielding
\begin{equation}
 {\rho}^{(2)}_{ 22}(t)=:W_2.
\end{equation}
This element describes the probability to find a dot with two
electrons. \\

\textbf{N = 3}. In this case we have again four eigenstates for
the system, whereas the two {\em odd} ones build the degenerate
groundstate and the two {\em even} ones are excited. In the
low-bias-regime we only need to deal with the 2 by 2 matrix
involving the three-particle groundstates
\begin{equation}
\left(\begin{array}{cc}
    {\rho}^{(3)}_{ 3o\uparrow 3o\uparrow} & {\rho}^{(3)}_{ 3o\uparrow 3o\downarrow}\\
     {\rho}^{(3)}_{ 3o\downarrow 3o\uparrow} & {\rho}^{(3)}_{ 3o\downarrow 3o\downarrow}
\end{array}\right)
=:\left(\begin{array}{cc}
     W_{3\uparrow} & w_3e^{i\alpha_3}\\
     w_3e^{-i\alpha_3} & W_{3\downarrow}
\end{array}\right).
\end{equation}
The total occupation probability for three electrons is
\begin{center}
$W_3:=W_{3\uparrow}+W_{3\downarrow}.$
\end{center}
As for the case $N=1$,  the off-diagonal elements yield
information on the  average spin
$S^{(3)}_i=\frac{1}{2}Tr\left(\sigma^{Pauli}_i\hat{\rho}^{(
3)}_\odot(t)\right)$ in the system through the relations
\begin{eqnarray}\label{S3}
S^{(3)}_x&=&= w_3 \cos{\alpha_3},\quad
S^{(3)}_y = -w_3 \sin{\alpha_3},\\
S^{(3)}_z&=&\frac{1}{2}(W_{3\uparrow}-W_{3\downarrow}).
\end{eqnarray}

 \textbf{N = 4}. Finally, if the double quantum dot is completely filled with
four electrons we only have one non-degenerate state.
Correspondingly, there is only one relevant RDM matrix element,
\begin{equation}
 {\rho}^{(4)}_{{44}}(t)=:W_4,
\end{equation}
describing the probability to find four electrons in the system.
The total spin is $S=0$.

Hence, we see that in all of the four cases (\ref{resonances}) we
get a system of five equations with the five independent physical
quantities $W_N, W_{N+1}$ and $ S_x^{(i)}, S_y^{(i)}, S_z^{(i)} $
with $i=1$ or 3.

\subsection{The conductance formula}
We shall exemplarily present results for the resonant transition
$N=1 \leftrightarrow N=2$. For the other transitions similar
considerations apply.
 The quantity
of interest are  $W_1, W_{2}, S^{(1)}_x, S^{(1)}_y, S^{(1)}_z $,
related through Eqs. (\ref{S1}), (\ref{S3}) to the
 density matrix elements of $\hat\rho^{(1)}_{\odot}. $
From Eqs. (\ref{Gl N}), (\ref{Gl N+1}), and the table \ref{tab:12}
of the appendix we finally obtain with $W_1=1 - W_2$,
\begin{eqnarray}
\label{masterW-S}\nonumber
\dot{W}_1&=&-\frac{\pi}{\hbar}\sum_{\alpha=s,d}\vert{t_{\alpha}}\vert^{2} k_+^2 \\
&& \Big( 2 F^{+}_{\alpha\downarrow\downarrow}(\mu_2)  W_1  - 4
F^{-}_{\alpha\downarrow\downarrow}(\mu_2)
 W_2 \nonumber \\
&-&4  F^{+}_{\alpha\uparrow\downarrow}(\mu_2) \vec{S}^{(1)} \cdot
\vec{m}_\alpha \Big)\;,
\\
\dot{\vec{S}}^{(1)}&=&-\frac{\pi}{\hbar}\sum_{\alpha=s,d}\vert{t_{\alpha}}\vert^{2}k_+^2
  \Big[2 F^{+}_{\alpha\uparrow\uparrow}(\mu_2) \vec{S}^{(1)}
 \nonumber \\
 &-&\Big(F^{+}_{\alpha\uparrow\downarrow}(\mu_2)W_1 -  2
 F^{-}_{\alpha\uparrow\downarrow}(\mu_2)W_2\Big)\vec{m}_\alpha \nonumber \\
   &+&\frac{2}{\pi k_+^2} \mathcal{P}_{\alpha}(\mu_{1},\left\lbrace
E_{2}\right\rbrace -E_1^{(0)})
\vec{m}_\alpha\times\vec{S}^{(1)}\Big].
\end{eqnarray}
 We have introduced the notation
$ 4 k_\pm^2:=(\alpha_0\pm\beta_0)^2$. All the non-vanishing
principle-value factors $P^{\pm}_{\alpha\uparrow\downarrow}$ and
the reflection-parameter $R_{\alpha\uparrow\downarrow}$ have been
merged to the compact form
\begin{eqnarray}
\lefteqn{\mathcal{P}_{\alpha}(\mu_{1},\left\lbrace
E_{2}\right\rbrace
-E_1^{(0)}):=-\frac{1}{2}[P^{-}_{\alpha\downarrow\uparrow}(\mu_1)+R_{\alpha\downarrow\uparrow}]}
\nonumber\\
&&-k_+^2 P^{+}_{\alpha\uparrow\downarrow}(\mu_2)+\frac{1}{4}P^{+}_{\alpha\uparrow\downarrow}(\varepsilon_{2^{\prime}}-\varepsilon_{1e})-\\
&&\hspace{0cm}-\frac{1}{4}P^{+}_{\alpha\uparrow\downarrow}(\varepsilon_{2^{\prime\prime}}-\varepsilon_{1e})-k_-^2
P^{+}_{\alpha\uparrow\downarrow}(\varepsilon_{2^{\prime\prime\prime}}-\varepsilon_{1e})\;,
\nonumber
\end{eqnarray}
where we introduced the chemical potential
$\mu_{N+1}=E_{N+1}^{(0)}-E_{N}^{(0)}$ and  $\left\lbrace E_2
\right\rbrace $ denotes the four different two particle energies.
We notice that the set of coupled Eqs. (\ref{masterW-S}) for the
evolution of the populations and of the spin accumulation has a
similar structure to that reported in
\cite{König03,Wetzels05,Koller07} for a single level quantum dot,
a metallic island and a single-walled carbon nanotube,
respectively. Some prefactors and the argument of the principal
part terms, however, are DD specific. In particular, as in
\cite{Wetzels05,Koller07}, we clearly identify a spin precession
term originating from the combined action of the reflection at the
interface and the interaction. The associated effective exchange
splitting is $\gamma B_1$, with $\gamma=-g\mu_B$ being the
gyromagnetic ratio, and
\begin{equation}
\vec{B}_{1}:=\frac{2}{\gamma}\sum_\alpha |t_\alpha|^2
\mathcal{P}_{\alpha}\vec{m}_\alpha \end{equation} being the corresponding effective exchange field. We focus now on
the stationary limit. In the absence of the precession term the
spin accumulation has only a $S_y^{(1)}$ component since, due to
our particular choice of the spin quantization axis, $S_x^{(1)}=0$
holds. The exchange field tilts the accumulated spin out of the
magnetizations' plane and gives rise to a nonzero $S_z^{(1)}$
component \emph{proportional} to $B_{1}$ and $S_y^{(1)}$.
%

\begin{figure}
 \hspace{-1.2cm}\includegraphics[width=7cm]{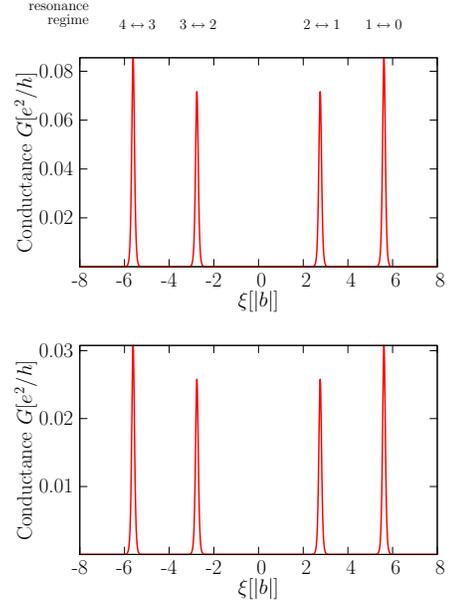}
 \caption{(Color online) Conductance at low bias for the parallel case $\Theta=0$ (top) and
 antiparallel case $\Theta=\pi$ (bottom).
 Notice the different peak heights and the mirror symmetry with respect to the
 half filling value $\xi=0$.  The conductance in the antiparallel configuration is always smaller
 than  in the parallel one. }
 \label{lowgate0}
\end{figure}

\begin{figure}
\includegraphics[width=7.5cm]{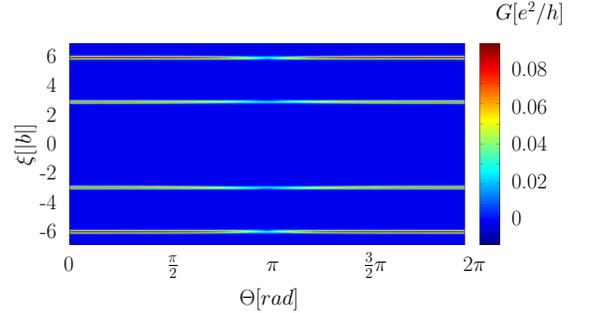}
\caption{(Color online) Conductance as a function of the polarization angle and
of  the gate voltage. The minimal conductance peaks occur as
expected at $\Theta=\pi$  }
 \label{lowgatephi_smallTemp}
\end{figure}

To get further insight in the spin-dynamics we observe that, since
we are looking at the low voltage regime, we can linearize the
Fermi function $f_\alpha$ in the bias voltage, i.e.,
\begin{equation} f_{\alpha}(\xi)=(1+e^{\beta(\xi+eV_{\alpha})})^{-1}\approx
f(\xi)(1-f(-\xi)e \beta V_{\alpha}). \end{equation} Introducing
the polarization of the contacts
\begin{equation*}
p_{\alpha}(\xi):=\frac{D_{\alpha\uparrow_{\alpha}}(\xi)-D_{\alpha\downarrow_{\alpha}}(\xi)}{D_{\alpha\uparrow_{\alpha}}
(\xi)+D_{\alpha\downarrow_{\alpha}}(\xi)}
\end{equation*}
we can express the
$F^{\pm}_{\alpha\sigma_\odot\sigma_\odot^{\prime}}$ factors as
\begin{eqnarray*}
F^{\pm}_{\alpha\uparrow\uparrow}(\xi)&\approx&\frac{1}{2}D_\alpha
(\xi) f(\pm\xi)\left( 1\mp f(\mp\xi) e\beta V_\alpha\right)\;,
 \\
F^{\pm}_{\alpha\uparrow\downarrow}(\xi)&=&
p_{\alpha}(\xi)F^{\pm}_{\alpha\uparrow\uparrow}(\xi) \;,
\end{eqnarray*}
where
$D_\alpha=D_{\alpha\uparrow_{\alpha}}+D_{\alpha\downarrow_\alpha}$.
It is also sufficient for our calculations to regard the density
of states as a constant quantity, $D_\alpha(\xi)=D_\alpha$.
Consequently the polarization is also constant,
$p_\alpha(\xi)=p_\alpha$. Finally, we focus in the following on
the symmetric case where both leads have the same properties,
which in  particular means that tunneling elements, polarizations,
density of states and reflection amplitude are equal:
\begin{eqnarray} t_1&=&t_2:=t,\quad p_1=p_2:=p,\nonumber \\
 D_1&=&D_2:=D,\quad R_{1\sigma_\odot -\sigma_\odot}=R_{2\sigma_\odot -\sigma_\odot}:=R\,.
\end{eqnarray} Upon introducing the linewidth
$\Gamma=\frac{2\pi}{\hbar}D\vert t \vert^2$
the conductance $G_{12}=I_{12}/V_{bias}$  for the resonant regime
$N=1\leftrightarrow N=2$ reads
\begin{eqnarray}
\label{I12} \lefteqn{G_{12}(\Theta)=\frac{\Gamma}{2}e^2 \beta
k_+^2 \frac{f(\mu_2)f(-\mu_2)}{f(-\mu_2)+1}}
\\&& \left(
1-\frac{p^2\sin^2(\frac{\Theta}{2})}{1+[B_1/f(\mu_2)2 \Gamma
k_+^2]^2
\cos^2(\frac{\Theta}{2})}\right).\nonumber
\end{eqnarray}
  Similarly we find for an arbitrary
resonance ($i=0,1,2,3$)
\begin{eqnarray}\label{I01}
\lefteqn{G_{i i+1}(\Theta)=\frac{\Gamma}{2}e^2 \beta |\langle
i+1|d^\dagger|i\rangle|^2
\frac{f(\mu_{i+1})f(-\mu_{i+1})}{1+f((-1)^i\mu_{i+1})}}
\\&& \hspace{-0.5cm}\left(
1-\frac{p^2 \sin^2(\frac{\Theta}{2})}{1+
[B_{i+1}/f((-1)^{i+1}\mu_{i+1})2\Gamma  |\langle
i+1|d^\dagger|i\rangle|^2]^2
\cos^2(\frac{\Theta}{2})}\right)\nonumber
\end{eqnarray}
%
%
where  $|\langle i+1|d^\dagger|i\rangle|$ is a shortcut notation
for the non vanishing matrix elements  $|\langle  E^{(0)}_{i+1}
i+1|d_{\alpha\odot}^\dagger| E^{(0)}_{i} i\rangle|$ calculated in
the tables of Appendix II. It holds $|\langle
1|d_{\alpha\odot}^\dagger| 0\rangle|=|\langle
4|d_{\alpha\odot}^\dagger| 3\rangle|=1/\sqrt{2}$, and $|\langle
2|d_{\alpha\odot}^\dagger| 1\rangle|=|\langle
3|d_{\alpha\odot}^\dagger| 2\rangle|=k_+$.
 Moreover, we
gathered together the principal part contributions and the ones
coming from the reflection Hamiltonian in the effective magnetic
fields
\begin{eqnarray*}
\vec{B}_2&=&\vec{B}_1,\\
\vec{B}_3&=&\vec{B}_4:=\frac{2}{\gamma}\sum_\alpha
|t_\alpha|^2\mathcal{P^\prime_\alpha}(\mu_4,E^{(0)}_3-\left\lbrace
E_2\right\rbrace)\vec{m}_\alpha\;.
\end{eqnarray*} The latter are defined in terms of the function
\begin{eqnarray*}
\lefteqn{\mathcal{P}^\prime_{\alpha}(\mu_{4},E^{(0)}_{3}-\left\lbrace
E_2\right\rbrace
):=-\frac{1}{2}[P^{+}_{\alpha\downarrow\uparrow}(\mu_4)+R_{\alpha\downarrow\uparrow}]}
\\
&&-k_+^2
P^{-}_{\alpha\uparrow\downarrow}(\mu_3)+\frac{1}{4}P^{-}_{\alpha\uparrow\downarrow}
(\varepsilon_{3o}-\varepsilon_{2^{\prime}})-\\
&&\hspace{0cm}-\frac{1}{4}P^{-}_{\alpha\uparrow\downarrow}(\varepsilon_{3o}-\varepsilon_{2^{\prime\prime}})-k_-^2
P^{-}_{\alpha\uparrow\downarrow}(\varepsilon_{3o}-\varepsilon_{2^{\prime\prime\prime}}).
\nonumber
\end{eqnarray*}
\begin{figure}
\hskip -0.7cm\includegraphics[width=8.8cm]{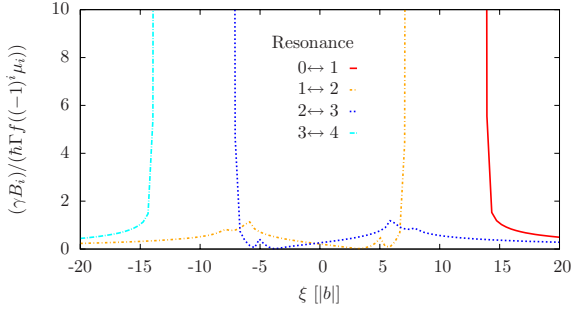}\vskip -0.6cm
 \caption{(Color online) Gate voltage dependence of the factors $B_i^2/f^2\Gamma^2$
 entering the conductance formula (\ref{I01}). Notice the mirror symmetry of the $0\leftrightarrow 1$ with the
 $3\leftrightarrow 4$ curve and of the $1\leftrightarrow 2$ with the $2\leftrightarrow 3$ one. }
 \label{picprincfermi}
 \end{figure}
 Moreover, a closer look to Eq. (\ref{I01})  shows that its angular dependence is strongly coupled to the square of the
ratio $(\gamma B_i)/(\hbar\Gamma f((-1)^i\mu_i))$, which is the effective exchange splitting, rescaled by the coupling and the Fermi function.
The ratio occurs in the denominators, and its value depends on the gate voltage. As the change of $B_i$ under variation of the gate voltage is comparatively small, the factor dominating the gate voltage evolution is the Fermi function. This accounts for the population of the dot: only if a nonzero spin is present (i.e. odd filling: one or three electrons), the effective magnetic field can have an influence. That is why correspondingly the renormalized effective exchange splitting vanishes for even fillings, namely below the $0\leftrightarrow1$ and $2\leftrightarrow3$, respectively above the $1\leftrightarrow2$ and $3\leftrightarrow4$ resonances. The curves belonging to the resonances involving the half filling do not go immediately to zero but show a more complex behavior with some small intermediate peaks due to the influence of the various excited states present for a two-electron population of the dot. This can nicely be seen from figure \ref{picprincfermi} (remember that $V_{gate}\propto -\xi$), where the four different factors
$(\gamma B_i)^2/(f((-1)^{i}\mu_i)\Gamma)^2$ are plotted. As we expect $G_{01}(\xi)$ and $G_{34}(\xi)$,
respectively $G_{12}(\xi)$ and $G_{23}(\xi)$ are mirror
symmetric with respect to each other when the gate voltage is varied. This in
turn reflects the electron-hole symmetry of the DD Hamiltonian.
The parameters of the figures are chosen to be ($b<0$)
\begin{eqnarray} k_B T &=& 4\cdot 10^{-2}|b|, \quad \hbar\Gamma = 4\cdot 10^{-3}|b|,\nonumber \\
U &=& 6 |b| ,\quad V= 1.6 |b| ,
\end{eqnarray}
and  $ p=0.8$, $R=0.05D$. As expected, the peaks are mirror
symmetric with respect to the half-filling gate voltage. Notice
also the occurrence of different peak  heights, both in the
parallel as well as in the antiparallel case.   For both
polarizations the principal-part-terms entering Eq. (\ref{I01})
vanish, the spin accumulation is entirely in the magnetization
plane and the peak ratio is solely determined by the ratio of the
groundstate overlaps $2/k_+^2$.
For polarization angles $\Theta\neq 0,\pi$ the ratio is also
determined by the non-trivial angular and voltage dependence of
the effective exchange fields.
Finally, as expected from the conductance formulas (\ref{I01}),
 the conductance is
suppressed in the antiparallel compared to the parallel case.
 The four
conductance peaks are plotted as a function of the gate
voltage in Fig. \ref{lowgate0} 
for the polarization angles $\Theta=0$ and  $\Theta=\pi$, top and
bottom figures, respectively. These features of the conductance
are nicely captured by the color plot of Fig.
\ref{lowgatephi_smallTemp}, where numerical results for the
conductance plotted as a function of gate voltage and polarization
angle are shown.
%
%
The conductance suppression nearby $\Theta=\pi$  is clearly seen.

%
%
%
In  the following we analyze in detail the single resonance
transitions. Due to the mirror symmetry it is convenient to
investigate together the resonances $N =0 \leftrightarrow N = 1$,
$N =3 \leftrightarrow N = 4$ and $N = 1 \leftrightarrow N = 2$, $N
= 2 \leftrightarrow N = 3$.
 We use the convention that, for a fixed resonance,
 the parameter $\xi=0$ when $\mu_{N+1}=0$.

\subsubsection{Resonant regimes $N = 0 \leftrightarrow N = 1$ and $N = 3 \leftrightarrow N = 4$}
\begin{figure}
\hspace{-1.2cm}\includegraphics[width=8.0cm]{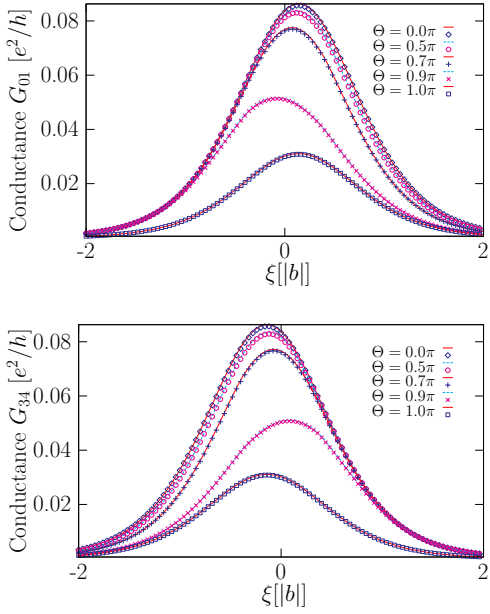}
 \caption{(Color online) The conductance $G_{01}(\xi)$ (upper figure) resp. $G_{34}(\xi))$ (lower figure)
 vs gate voltage for different polarization angles. The mirror
 symmetry
 of the conductance peaks for the $0\leftrightarrow1$ and $3\leftrightarrow4$ transitions is
 clearly observed.
 Notice the
 excellent agreement between the prediction of the analytical formula Eq. (\ref{I01})
 (continuous lines)
 and the results of a  numerical integration of Eqs. (\ref{master}) with (\ref{current}) (symbols).}
 \label{0134gate}
 \end{figure}

The expected mirror symmetry of  $G_{01}$ and $G_{34}$  is shown
in Figure \ref{0134gate}, where the conductance peaks are plotted
for different polarization angles $\Theta$ of the contacts. Notice
that the analytical expressions (\ref{I01})  (continuous lines)
perfectly match the results obtained from a numerical integration
of the master equation (\ref{master}) with the current formula
(\ref{current}). We also can see that the maxima of the
conductance decrease with $\Theta$ growing up to $\pi$.
It can be shown that the peaks for $\Theta=0$ and $\Theta=\pi$ lie
at the same value of $\xi$, because the effective fields $B_i$
exactly vanish due to trigonometrical prefactors. In other words,
virtual processes captured in the effective fields $B_i$ do not
play a role in the collinear case. For noncollinear configurations, however, the peak maxima are shifted towards the gate voltages where an odd population of the dot dominates, because there the effective exchange field can act on the accumulating spin and makes it precess, which eases tunneling out. These findings are in agreement with
results obtained for a single-level quantum dot \cite{König03}, a
metallic island \cite{Wetzels05} and carbon nanotubes
\cite{Koller07}. 
%
\begin{figure}
\hspace{-0.6cm}\includegraphics[width=7.5cm]{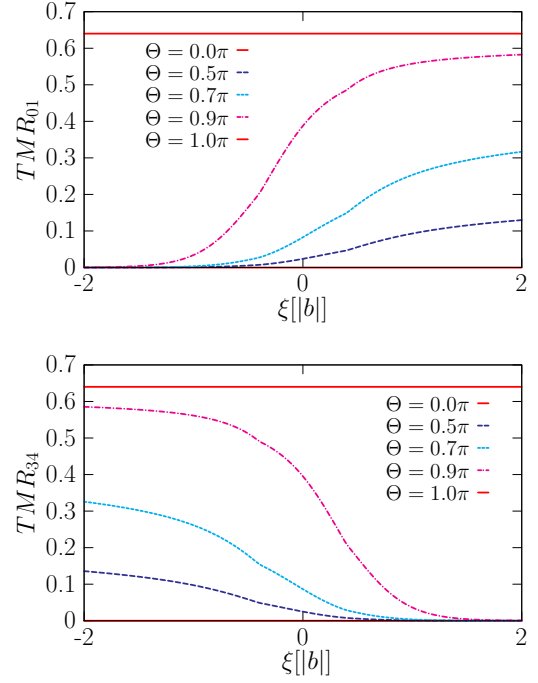}
 \caption{(Color online) Tunneling magnetoresistance (TMR) for the transitions $0\leftrightarrow 1$
  (upper figure) and  $3\leftrightarrow 4$ (lower figure)  vs gate
  voltage. The TMR is always positive and is independent of gate
  voltage for collinear lead magnetizations only, $\Theta=0$ and
  $\Theta=\pi$.}
 \label{TMR0134}
 \end{figure}
To quantify the relative magnitude of the current for a given
polarization angle $\Theta$ with respect to the case $\Theta=0 $
we introduce the angle-dependent tunneling magnetoresistance (TMR)
as
\begin{equation*}
TMR_{N\,N+1}(\Theta,\xi)=1-\frac{G_{N,N+1}(\Theta,\xi)}{G_{N,N+1}(0,\xi)}.
\end{equation*}
For the transition $0\leftrightarrow 1$ it reads
\begin{equation}\label{eqntmr}
TMR_{01}
=\frac{p^2 \sin^2(\frac{\Theta}{2})}{1+[B_1^2/f^2(-\mu_1)\Gamma^2]
\cos^2(\frac{\Theta}{2})}\;.
\end{equation}
\begin{figure}
\hspace{-0.6cm}\includegraphics[width=7.5cm]{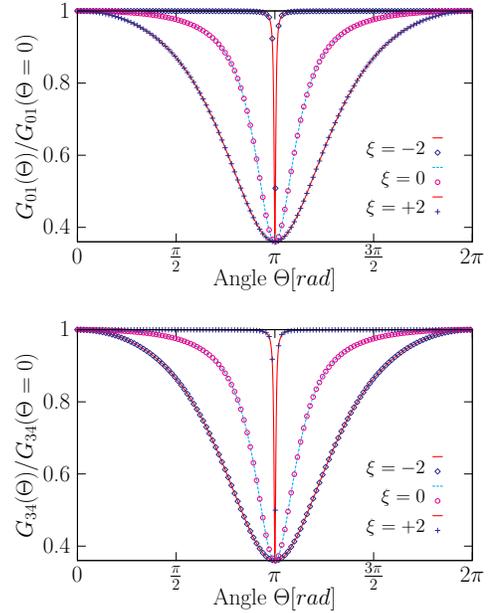}
 \caption{(Color online) $G_{01}(\Theta)/G_{01}(0)$ (upper figure) resp. $G_{34}(\Theta)/G_{34}(0)$ (lower figure)
 vs polarization angle. For all the three chosen values of the
 gate voltage the curve display an absolute minimum at
 $\Theta=\pi$. Notice the
  overall agreement  of the analytical predictions (\ref{I01})
  given by the continuous curves with outcomes of a numerical solution of the master
  equation, Eq. (\ref{master}), together with (\ref{current})(symbols).}
 \label{0134phi}
 \end{figure}
\begin{figure}[h]
 \hspace{-0.6cm}\includegraphics[width=7.5cm]{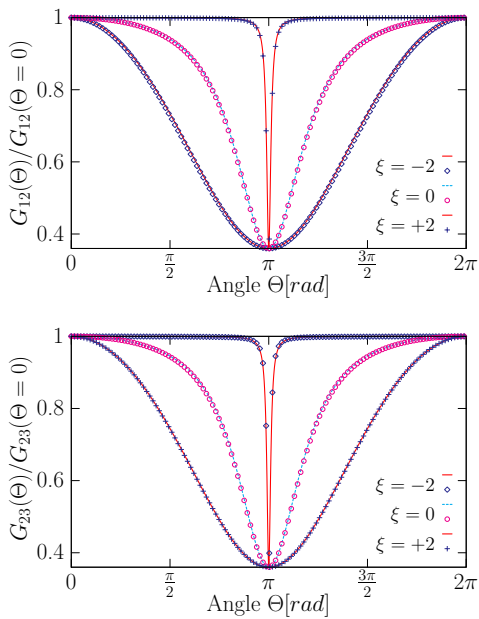}\\
 \caption{(Color online) $G_{12}(\Theta)/G_{12}(0)$ (upper figure) resp. $G_{23}(\Theta)/G_{23}(0)$ (lower figure)
 vs polarization angle. Notice the overall agreement of the analytical predictions (\ref{I12}) and (\ref{I01}),
  continuous lines, with the
  data (symbols) coming from a
 numerical  solutions of the equations for the reduced density matrix.}
 \label{1223phi}\end{figure}
Hence, the TMR vanishes for $\Theta=0$ and takes the  constant
value
\begin{equation*}
TMR_{01}(\pi,\xi)=p^2
\end{equation*}
at  $\Theta=\pi$. For the remaining polarization angles,
$\Theta\neq0$ and $\Theta\neq\pi$, the TMR is gate-voltage
dependent and positive.
 The behavior of the TMR as a function of the
gate voltage is shown in figure \ref{TMR0134}. To understand the
gate voltage dependence of the TMR  at noncollinear angles we have to
remember that the dot is depleted with raising $\xi$. For the
transition $0 \leftrightarrow 1$, this means that at positive $\xi$,
the dot is predominantly empty, so that an electron which
enters the dot also fast leaves it. In this situation the TMR is
finite and its value depends in a complicated way on the amplitude
of the exchange field. At negative $\xi$, the DD is predominantly
occupied with an electron which can now interact with the exchange field,
which makes the spin precess and thus eases tunneling out of the dot. Consequently,
$G_{N N+1}(\Theta,\xi)\approx G_{N N+1}(0,\xi)$ and the TMR vanishes.
Finally, figure \ref{0134phi} illustrates the angular dependence
of the normalized conductance for three different values of the
gate voltage. We detect a common absolute minimum for the
conductance at $\Theta=\pi$, i.e., transport is weakened in the
antiparallel case. The width of the curves is dependent on the renormalized
effective exchange $(\gamma B_i)/(\Gamma f)$. The larger its value the narrower
get the curves, because the spin precession can equilibrate the accumulated spin for
all angles but $\Theta=\pi$. Notice again the equivalence of the curves
belonging to $\xi = \pm 2  |b |$ for the $1\leftrightarrow 2$ resonance to the
curves with $\xi = \mp 2 |b | $ for the $3\leftrightarrow 4$ resonance.

\subsubsection{Resonant regimes $N = 1 \leftrightarrow N = 2$
and $N = 2 \leftrightarrow N = 3$}

For the resonant transitions $1 \leftrightarrow 2$ and $2
\leftrightarrow 3$ qualitatively analogous results as for the $0
\leftrightarrow 1$ and $3 \leftrightarrow 4$ transitions are
found. Thus, exemplarily we only show the angular dependence of
the normalized conductance in  Fig. \ref{1223phi}, showing the
expected absolute conductance minimum at $\Theta=0$.
%
%
%

\section{Nonlinear transport}\label{high bias}
\begin{figure}
 \includegraphics[width=8.2cm]{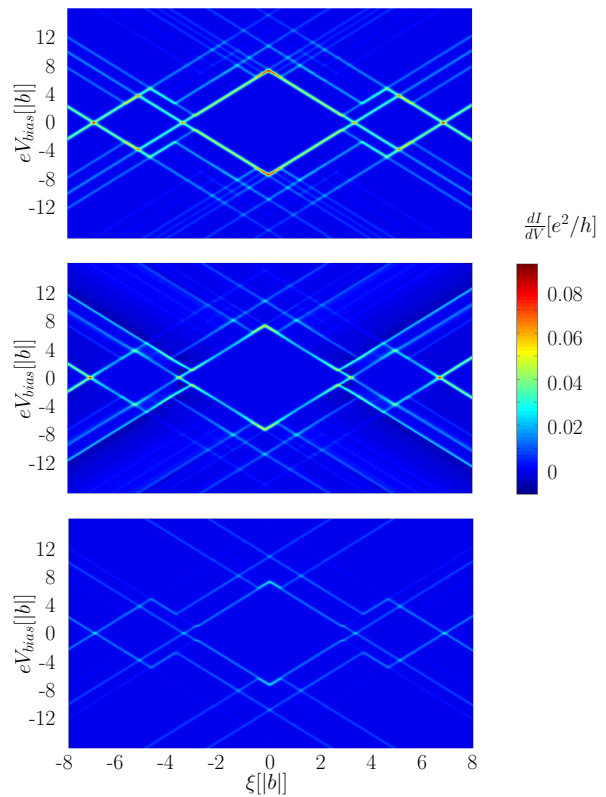}
 \caption{(Color online) Differential conductance
 $\frac{dI}{dV}$  for the parallel $\Theta=0$ (top), perpendicular $\Theta=\pi/2$ (middle) and antiparallel
  $\Theta =\pi$ (bottom) configurations. The two half
  diamonds and three diamonds regions correspond to bias and gate voltage values where transport is Coulomb blocked.
 The excitation lines, where  excited states contribute to resonant transport, are clearly visible in all of the three cases.
 However, a negative differential conductance is observed in the perpendicular case, while
 some excitation lines are absent in the antiparallel configuration. }
 \label{vvdi0}
\end{figure}
%
In this section we present the numerical results, deduced from the
general master equation (\ref{master}) combined with the current
formula (\ref{current}).
\begin{figure}
\hspace{-1.2cm}\includegraphics[width=6.5cm]{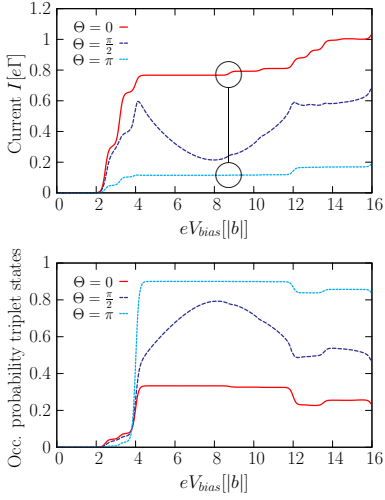}
\caption{(Color online) Current (top) and triplet occupation (bottom) for the
two collinear ($\Theta=0$,$\Theta=\pi$) cases and the
perpendicular case ($\Theta=\frac{\pi}{2}$) at a fixed gate
voltage $\xi = 4 |b|$. Notice the occurrence of a pronounced
negative differential conductance feature for perpendicular
polarization $\Theta = \pi/2$. } \label{NDCcurrent}
\end{figure}
%
We show the differential conductance $\frac{dI}{dV}(\xi,V_{bias})$
for the three distinct angles
$\Theta=0$, $\Theta=\frac{\pi}{2}$ and $\Theta=\pi$, see figure
\ref{vvdi0}, top, middle and bottom,
respectively.
The results confirm the electron-hole-symmetry and the symmetry
upon   bias voltage inversion $I(\xi,V_{bias})=-I(\xi,-V_{bias})$.
In all of the three cases we can nicely see the expected three
closed and the two half-open diamonds, where the current is
blocked and the electronic number of the double-dot-system stays
constant. At higher bias voltages the contribution of excited
states
 is manifested in the appearance of several  excitation lines.
One clearly sees that transition lines present in the parallel
case are {\em absent} in the antiparallel case.  Moreover, in the
case of noncollinear polarization, $\Theta =\pi /2$, negative
differential conductance (NDC) is observed.\\
In the following, we want not only to explain the origin of these
two features, but alongside also give another example for
spin-blockade effects, which play a decisive role in the DD physics.\\
As a starting point, we plot in figure \ref{NDCcurrent} (top) the
current through the system for the three different angles
$\Theta=\left\lbrace 0, \frac{\pi}{2}, \pi\right\rbrace $ at a
fixed gate voltage  $\xi=4 |b|$ and positive bias voltages. We
recognize, that for $eV_{Bias}<2.4 |b|\,$ the current is
Coulomb-blocked in all of the three cases. In this configuration
exactly one electron stays in the double-dot. From about
$eV_{Bias}\geq 2.4 |b|\,$ the channel where the groundstate energies
$\mu_1$ and $\mu_2$ are degenerate opens ($| 1e \sigma\rangle
\leftrightarrow |2 \rangle$ transition) and current begins to
flow. With increasing bias more and more transport channels become
energetically favorable. In particular, for all the polarization
angles $\Theta$ we observe two consecutive steps corresponding to
the transitions $|0 \rangle \leftrightarrow |1e \sigma\rangle$ and
$|1e \sigma \rangle \leftrightarrow |2'\rangle$. The
latter, occurring at about $eV_{bias}=4|b|$, involves the excited
two-particle triplet states $|2'(S_z)\rangle$.\\
The next excitation step, indicated with a
circle in Fig. \ref{NDCcurrent} (top), belongs to the
transition $|1o \sigma \rangle \leftrightarrow |2''\rangle$.
The associated line is missing for the antiparallel configuration, as well as 
the lines corresponding to $|1o \sigma \rangle
\leftrightarrow |2'''\rangle$; $|1e \sigma\rangle \leftrightarrow
|2''\rangle$; $|2\rangle \leftrightarrow |3o \sigma \rangle$; $|1e
\sigma\rangle \leftrightarrow |2''' \rangle$. Crucially, in all of
these transitions a two-particle state with \textit{total spin zero} is
involved. In order to explain the absence of these lines, let us e.g. focus on
the first missing step corresponding to the $|1o \sigma \rangle \leftrightarrow
|2''\rangle$ resonance. In the parallel case (say both contacts polarized
spin-up) there is always an open channel corresponding to the situation in
which the spin in the DD is antiparallel to that in the leads (i.e. $|1o -
\rangle$). In the antiparallel case (say source polarized spin-up, drain
polarized spin-down) originally a spin-down might be present in the dot. An
electron which enters the DD from the source must then be spin-up (in order to
form the state $|2''\rangle$), but as the drain is down-polarized, it will be
the spin-down electron which leaves the DD, which corresponds to a spin flip.
Now the presence of a spin-up electron in the DD prevents a majority (another
spin-up) electron from the source to enter the DD, such that we end up in a
blocking state.
The transition is hence forbidden.

A similar, yet different spin-blockade effect determines the occupation probabilities for
the triplet state, Fig. \ref{NDCcurrent} (bottom). Naturally, for all angles the probability to be in the triplet state increases
above the resonance at $eV_{bias}=4|b|$, but interestingly, such
probability is \emph{largest} in the antiparallel case. This is
due to the fact that a majority spin in the parallel configuration (spin-up) can be easily transmitted
through the DD via the triplet states $|2'(1)\rangle$ or
$|2'(0)\rangle$.
 In the antiparallel case, however, a blocking state establishes (say again source polarized spin-up, drain polarized spin-down).
Let initially a spin-down electron be present on the DD. From the source electrode, most likely 
a majority electron polarized spin-up electron will enter the dot. Now, just as in the previous case,
the consecutive tunneling event will cause a spin flip in the DD, because the spin-down electron (majority electron of the drain) will leave the dot. 
So the DD is finally in a spin-up state, and once the next majority spin-up electron from the source enters, the DD ends up in the triplet state
$|2'(+1)\rangle$ and will remain there for a long time
due to the fact that the majority spins in the drain are down-polarized. Hence the triplet state $|2'(+1)\rangle$ acts as a
trapping state.

Notice that the two distinct spin-blockade effects are
different from the Pauli spin-blockade discussed in the DD
literature \cite{Ono02,Liu05,Johnson05,Fransson06}. Moreover, the
second effect, relying on the existence of degenerate triplet states, is also
different from the spin-blockade found in Ref. \cite{König03} for
a single level quantum dot.

Finally, let us turn to the negative differential conductance, which occurs for
noncollinearly polarized leads (see the dashed blue lines in figure \ref{NDCcurrent}), 
and which we find to become more evident for higher polarizations (not shown).
Neglecting the exchange field, we would just expect the magnitude of the current for the noncollinear
polarizations to lie somewhere in between the values for the parallel and the antiparallel current, because
the noncollinear polarization could in principle be rewritten as a linear combination of the parallel
and the antiparallel configuration. Now the effect of the exchange is to cause precession and therewith
equilibration of the accumulating spin, which corresponds to shifting the balance in favor of the parallel
configuration, i.e. enhancing the current. The decisive point is that the exchange field is not only gate,
but also bias voltage dependent and reaches a minimum around $eV_{Bias}\approx 8 |b|$. This explains the
decreasing of the current up to this point. Afterwards the influence of the spin precession
regains weight. The same consideration applies for the other NDC regions observed in Fig. \ref{vvdi0}, e.g. in
the gate voltage region $\xi\approx 2 |b|$ involving the $N=0
\leftrightarrow N=1$ transition, as described in Ref. \cite{König03}.

\section{The effects of an  external magnetic field }\label{magnetic}
In this section we wish to discuss the qualitative changes brought
by an external magnetic field applied to the DD. Specifically, the
magnetic field is assumed to be parallel to the magnetization
direction of the drain. For simplicity we focus on the
experimental standard case of parallel and antiparallel lead
polarization and of low bias voltages. Then, the magnetic field
causes an energy shift $\mp E_{Zeeman}$ depending on whether the
electron spin is parallel or antiparallel, respectively, to it.
For collinear polarization angles the principal part contributions
vanish, and the equations for the RDM are easily obtained. We
report exemplarily results for the transitions $0\leftrightarrow
1$ and $1\leftrightarrow 2$.
\begin{figure}[ht]
\vspace{0.3cm}
\hspace{-1.2cm}\includegraphics[width=7cm]{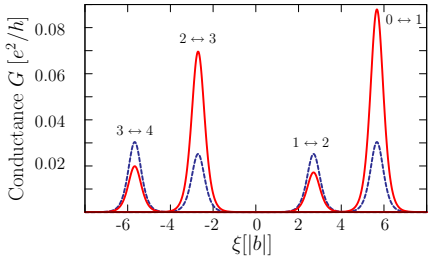}
\caption{(Color online) Conductance vs gate voltage for parallel (continuous
line) and antiparallel (dashed lines) contact configuration and
Zeeman splitting $E_{Zeeman}=0.05 |b|$. The magnetic field breaks
the mirror symmetry with respect to the gate voltage in the {\em
parallel} configuration. \label{magnetcurrent}}
\vspace{0.3cm}
\end{figure}
Let us then consider the  parameter regime nearby the
$0\leftrightarrow 1$ resonance, and  setup a system of three
equations with three unknown variables $W_0, W_{1\uparrow}$ and
$W_{1\downarrow}$. The first equation corresponds to the
normalization condition $ W_{1\uparrow}+W_{1\downarrow}+W_{0}=1 $.
The remaining equations are the equations of motion for
$W_{1\uparrow/\downarrow}$ which  can be written as
\begin{eqnarray}
\dot{W}_{1\uparrow}&=&-\frac{\pi}{\hbar}\sum_{\alpha=s,d}\vert
t_\alpha\vert^2\left[ F^-_{\alpha\uparrow}(\mu_{1\uparrow})
W_{1\uparrow} \right.  \nonumber\\
&-& \left. F^+_{\alpha\uparrow}(\mu_{1\uparrow})W_0 \right],
\end{eqnarray}
\begin{eqnarray}
\dot{W}_{1\downarrow}&=&-\frac{\pi}{\hbar}\sum_{\alpha=s,d}\vert
t^\alpha\vert^2\left[
F^-_{\alpha\downarrow}(\mu_{1\downarrow})W_{1\downarrow}
\right. \nonumber\\
 &-& \left.
F^+_{\alpha\downarrow}(\mu_{1\downarrow})W_0 \right],
\end{eqnarray}
where
\begin{equation}
\mu_{1\uparrow/\downarrow}=\mu_{1}\mp E_{Zeeman},
\end{equation}
and $F^{\pm}_{\alpha\sigma_\odot}(E)=D_{\alpha\sigma_\odot}f^\pm
(E)$. For the collinear case is $D_{\alpha\sigma_\odot}=D_{\alpha
\pm_\alpha}$ if $\sigma_\odot=\uparrow/\downarrow$ in the parallel
case. On the other hand  $D_{s\sigma_\odot}=D_{s \mp_s}$ and
$D_{d\sigma_\odot}=D_{d \pm_d}$, if
$\sigma_\odot=\uparrow/\downarrow$, in the antiparallel case.

\begin{figure}[h]
\vspace{0.2cm}
\hspace{-0.6cm}\includegraphics[width=6cm]{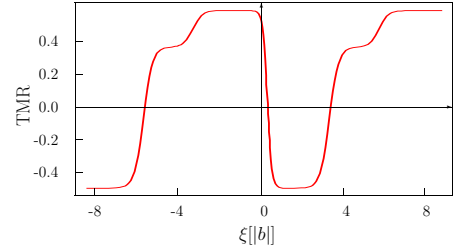} \caption{(Color online) Tunneling
magnetoresistance (TMR) vs gate voltage in the presence of an
external magnetic field. In contrast to the zero field case, the
TMR can become negative in the vicinity of the $2\leftrightarrow
3$ and $3\leftrightarrow4$ resonances. \label{magnetTMR}}
\end{figure}
Upon considering symmetric contacts
  ($t_1=t_2=t$, $D_1=D_2=D$) we find in the parallel case
\begin{eqnarray}
 \lefteqn{G_{01}(\Theta=0)=
\frac{\Gamma e^2}{8\beta}f(-\mu_{1\uparrow})f(-\mu_{1\downarrow})}\\
&\times&\frac{ p\left(
f(\mu_{1\uparrow})-f(\mu_{1\downarrow})\right)
+f(\mu_{1\uparrow})+f(\mu_{1\downarrow})}{f(-\mu_{1\uparrow})f(\mu_{1\downarrow})+f(-\mu_{1\uparrow})f(-\mu_{1\downarrow})+f(\mu_{1\uparrow})f(-\mu_{1\downarrow})}
.\nonumber
\end{eqnarray}
For the antiparallel case we obtain
\begin{eqnarray}
 \lefteqn{G_{01}(\Theta=\pi)=G_{01}(\Theta=0)}\\
&&\times\frac{1-p^2\left(
f(\mu_{1\uparrow})+f(\mu_{1\downarrow})\right)
}{p\;[f(\mu_{1\uparrow})-f(\mu_{1\downarrow})]+f(\mu_{1\uparrow})+f(\mu_{1\downarrow})
} .\nonumber
\end{eqnarray}
Analogously we find for the $1\leftrightarrow 2$ transition
\begin{eqnarray}
\lefteqn{ G_{12}(\Theta=0)=\frac{\Gamma e^2 k_{+}^2}{2\beta}
f(\mu_{2\uparrow})f(\mu_{2\downarrow})}\\
&\times&\frac{ p\left(
f(-\mu_{2\downarrow})-f(-\mu_{2\uparrow})\right)
+f(-\mu_{2\uparrow})+f(-\mu_{2\downarrow})}{f(-\mu_{2\uparrow})f(\mu_{2\downarrow})+f(-\mu_{2\uparrow})f(-\mu_{2\downarrow})+f(\mu_{2\uparrow})f(-\mu_{2\downarrow})}
\nonumber ,
\end{eqnarray}
\begin{eqnarray}
\lefteqn{ G_{12}(\Theta=\pi)=G_{12}(\Theta=0)}\\
&\times& \frac{1 -p^2 (
f(-\mu_{2\uparrow})+f(-\mu_{2\downarrow}))}{p\;,[f(-\mu_{2\downarrow})-f(-\mu_{2\uparrow})]+f(-\mu_{2\uparrow})+f(-\mu_{2\downarrow})}
\nonumber .
\end{eqnarray}
The remaining resonances are calculated analogously. Figure
 \ref{magnetcurrent} shows the four conductance resonances for
 the parallel and antiparallel configurations.
 Strikingly,
the applied magnetic field breaks the symmetry between the
tunneling regimes $0\leftrightarrow1$ and $3\leftrightarrow4$, as
well as between the resonances
 $1\leftrightarrow2$ and $2\leftrightarrow3$ in case of {\em parallel} contact polarizations.
 The reason for this behavior is the following:
In the low bias regime transitions between groundstates dominate
transport. In particular, the magnetic field removes the spin
degeneracy of the states $|1e \sigma \rangle $ and of the states
$|3o \sigma \rangle $, such that states with spin aligned to the
external magnetic field are energetically favored. Therefore, the
transport electron in the tunneling regime $0\leftrightarrow 1$ is
a majority spin carrier. For the case $3\leftrightarrow 4$,
however, two of the three electrons of the groundstate $|3o+\rangle $ have spin-up, such that the fourth electron
which can be added to the DD has to be a minority spin carrier.
Therefore, the conductance gets diminished with respect to the
$0\leftrightarrow 1$ transition, and the mirror symmetry present
in the zero field case is broken. Analogously the broken symmetry
in the case of the transitions $1\leftrightarrow 2$ and
$2\leftrightarrow 3$ can be understood.
Correspondingly, the TMR can become negative for values of the
gate voltages around the $2\leftrightarrow 3$ and
$3\leftrightarrow4$ resonances. We observe that a negative TMR has
recently been predicted in \cite{Cottet06} for the case of a
single impurity Anderson model with orbital and spin degeneracy.
In that work a negative TMR arises due to the assumption that
multiple reflections at the interface cause spin-dependent energy
shifts. In our approach, however, where the contribution from the
reflection Hamiltonian  is treated to the lowest order, see
(\ref{href}), such spin-dependent energies are originating from
the magnetic field-induced Zeeman splitting.

\section{Conclusions}\label{conclusions}
In summary we have evaluated linear and nonlinear transport
through a double quantum dot (DD) coupled to polarized leads with
arbitrary polarization directions. Due to strong Coulomb
interactions  the DD operates as a single-electron transistor, a
F-SET, at low enough temperatures. A detailed analysis of the
current-voltage characteristics of the DD, and comparison with
results of previous studies on other F-SET systems with
noncollinear (a single level quantum dot \cite{König03}, a
metallic island \cite{Wetzels05},
 a carbon nanotube \cite{Koller07}), bring us to the
 identification of
 {\em universal } behaviors of a F-SET, i.e., a behavior shared
by {\em any} of those F-SETs, independent of the specific kind of
conductor is considered as the central system, as well as
system-specific features.\\
Universal is the presence of an
interfacial exchange field together with an interaction-induced
one, present  for noncollinear polarization only. These exchange
fields cause a precession of the accumulated spin on the dot and
therewith ease the tunneling out. This effect has various implications.
It determines e.g. the gate and angular dependence, as well as the height of single
conductance peaks and can yield negative differential conductance
features.  Another universal feature is the occurrence of a negative tunneling magnetoresistance -
even in the weak tunneling limit - if an external magnetic field
is applied.\\
Specific to the DD system are the following features:
In the low bias regime, the problem can be solved analytically and
for the tunneling regimes $0\leftrightarrow1$ and $1\leftrightarrow2$, respectively
$2\leftrightarrow3$ and $3\leftrightarrow4$, the system behaves equivalent to a single
level quantum dot, where the Coulomb blockade peaks are found to be mirror-symmetric with respect to the charge neutrality point.
This mirror symmetry reflects
the electron-hole-symmetries of a system and is therefore typical for the DD, as well as the ratio of the peak heights.
An external magnetic field lifts this symmetry and can cause a negative tunneling magnetoresistance. 
In the nonlinear bias regime, the presence of various excited states gives rise to interesting DD specific features. For example, a suppression of several
excitation lines for an antiparallel lead configuration originates from a spin-blockade effect. It occurs
because a trapping state is formed whenever
a transition involves a two-electron state with total spin zero.
A second spin-blockade effect we described involves the two-electron triplet
state.  The common mechanism of these two spin-blockades is the following: in
both cases, a tunneling event can only occur if initially the dot is populated
with an unpaired electron possessing the majority spin of the drain.  The
second step is that a majority electron of the source will enter, forming a
spin-zero state. Then the first electron can leave the dot, causing a
spin-flip.  If the triplet state is involved, the dot will be left in a
trapping state once a second majority electron from the source enters.
Otherwise, we are directly in a blocking state.
Finally, for noncollinear lead polarizations,
negative differential conductance can be observed.\\
All in all, due
to their universality and the multiplicity of their properties,
F-SET based on DD systems seem good candidates for future
magneto-electronic devices.

 \acknowledgements Financial support  under the DFG
programs SFB 689 and SPP 1243  is acknowledged.

\textheight20cm

\begin{widetext}
\vskip -0.28cm
\clearpage
\end{widetext}
\textheight235mm

\appendix

\section{Matrix elements of the dot operators \label{appendix2}}
The following tables show  all the possible matrix-elements
$\langle{N-1}\vert d_{\alpha\sigma_\odot}\vert N\rangle$ and
$\langle{N}\vert d^\dagger_{\alpha\sigma_\odot}\vert {N-1}\rangle$
with $\alpha=\left\lbrace 1,2\right\rbrace $ and
$\sigma_\odot=\left\lbrace \uparrow,\downarrow\right\rbrace $
which occur in the master equations (\ref{Gl N}) and (\ref{Gl
N+1}). Notice that we only need to illustrate either the matrix
elements $\langle{N-1}\vert d_{\alpha\sigma_\odot}\vert N\rangle$
or $\langle{N}\vert d^\dagger_{\alpha\sigma_\odot}\vert
{N-1}\rangle$, because they are  complex conjugated to each other.

%
\vspace{0.5cm}
\begin{table}[h]
    \centering
  \tiny\renewcommand{\arraystretch}{1.5}

        \begin{tabular}{|ccccc|}
            \hline
            &&$0 \leftrightarrow 1$&&\\
            \hline\hline
            &$\vert{1e \uparrow}\rangle$&$\vert{1e \downarrow}\rangle$&$\vert{1o \uparrow}\rangle$
            &$\vert{1o \downarrow}\rangle$\\
            \hline
            $d_{1\uparrow}:\bra{0}$&$\frac{1}{\sqrt{2}}$&$0$&$\frac{1}{\sqrt{2}}$&0\\
            \hline
            $d_{1\downarrow}:\bra{0}$&0&$\frac{1}{\sqrt{2}}$&0&$\frac{1}{\sqrt{2}}$\\
            \hline
            $d_{2\uparrow}:\bra{0}$&$\frac{1}{\sqrt{2}}$&0&$-\frac{1}{\sqrt{2}}$&0\\
            \hline
            $d_{2\downarrow}:\bra{0}$&0&$\frac{1}{\sqrt{2}}$&0&$-\frac{1}{\sqrt{2}}$\\
            \hline
        \end{tabular}
    \caption{Matrix elements for the transition $N=0\leftrightarrow N=1$  induced by the operators
    $d_{\alpha\uparrow}$ and $d_{\alpha\downarrow}$, $\alpha=1,2$. }
    \label{tab:10}
\vspace{1cm}

    \tiny\renewcommand{\arraystretch}{1.5}

        \begin{tabular}{|c|ccccc|}
            \hline
            &&&$1 \leftrightarrow 2$&&\\
            \hline\hline
            &&$\vert{1e \uparrow}\rangle$&$\vert{1e \downarrow}\rangle$&$\vert{1o \uparrow}\rangle$
            &$\vert{1o \downarrow}\rangle$\\
            \hline
            &$\bra{2g}$&$0$&$\frac{\alpha_0+\beta_0}{2}$&$0$&$\frac{-\alpha_0+\beta_0}{2}$\\
            \cline{2-6}
            &$\bra{2^\prime(+1)}$&$\frac{1}{\sqrt{2}}$&$0$&$-\frac{1}{\sqrt{2}}$&$0$\\
          \cline{2-6}
            $d^\dagger_{1\uparrow}:$&$\bra{2^\prime(0)}$&$0$&$\frac{1}{2}$&$0$&$-\frac{1}{2}$\\
            \cline{2-6}
            &$\bra{2^\prime(-1)}$&$0$&$0$&$0$&$0$\\
          \cline{2-6}
            &$\bra{2^{\prime\prime}}$&$0$&$\frac{1}{2}$&$0$&$\frac{1}{2}$\\
         \cline{2-6}
 &$\bra{2^{\prime\prime\prime}}$&$0$&$\frac{-\alpha_0+\beta_0}{2}$&$0$&$\frac{-\alpha_0-\beta_0}{2}$\\

      \hline\hline
            &$\bra{2g}$&$\frac{-\alpha_0-\beta_0}{2}$&$0$&$\frac{\alpha_0-\beta_0}{2}$&$0$\\
            \cline{2-6}
            &$\bra{2^\prime(+1)}$&$0$&$0$&$0$&$0$\\
          \cline{2-6}
            $d^\dagger_{1\downarrow}:$&$\bra{2^\prime(0)}$&$\frac{1}{2}$&$0$&$-\frac{1}{2}$&$0$\\
            \cline{2-6}
            &$\bra{2^\prime(-1)}$&$0$&$\frac{1}{\sqrt{2}}$&$0$&$-\frac{1}{\sqrt{2}}$\\
          \cline{2-6}
            &$\bra{2^{\prime\prime}}$&$-\frac{1}{2}$&$0$&$-\frac{1}{2}$&$0$\\
         \cline{2-6}
 &$\bra{2^{\prime\prime\prime}}$&$\frac{\alpha_0-\beta_0}{2}$&$0$&$\frac{\alpha_0+\beta_0}{2}$&$0$\\
            \hline\hline

                &$\bra{2g}$&$0$&$\frac{\alpha_0+\beta_0}{2}$&$0$&$\frac{\alpha_0-\beta_0}{2}$\\
            \cline{2-6}
            &$\bra{2^\prime(+1)}$&$-\frac{1}{\sqrt{2}}$&$0$&$-\frac{1}{\sqrt{2}}$&$0$\\
          \cline{2-6}
            $d^\dagger_{2\uparrow}:$&$\bra{2^\prime(0)}$&$0$&$-\frac{1}{2}$&$0$&$-\frac{1}{2}$\\
            \cline{2-6}
            &$\bra{2^\prime(-1)}$&$0$&$0$&$0$&$0$\\
          \cline{2-6}
            &$\bra{2^{\prime\prime}}$&$0$&$-\frac{1}{2}$&$0$&$\frac{1}{2}$\\
         \cline{2-6}
 &$\bra{2^{\prime\prime\prime}}$&$0$&$\frac{-\alpha_0+\beta_0}{2}$&$0$&$\frac{\alpha_0+\beta_0}{2}$\\

      \hline\hline
            &$\bra{2g}$&$\frac{-\alpha_0-\beta_0}{2}$&$0$&$\frac{-\alpha_0+\beta_0}{2}$&$0$\\
            \cline{2-6}
            &$\bra{2^\prime(+1)}$&$0$&$0$&$0$&$0$\\
          \cline{2-6}
            $d^\dagger_{2\downarrow}:$&$\bra{2^\prime(0)}$&$-\frac{1}{2}$&$0$&$-\frac{1}{2}$&$0$\\
            \cline{2-6}
            &$\bra{2^\prime(-1)}$&$0$&$-\frac{1}{\sqrt{2}}$&$0$&$-\frac{1}{\sqrt{2}}$\\
          \cline{2-6}
            &$\bra{2^{\prime\prime}}$&$\frac{1}{2}$&$0$&$-\frac{1}{2}$&$0$\\
         \cline{2-6}
 &$\bra{2^{\prime\prime\prime}}$&$\frac{\alpha_0-\beta_0}{2}$&$0$&$\frac{-\alpha_0-\beta_0}{2}$&$0$\\
            \hline
        \end{tabular}
    \caption{Matrix elements for the transition  $N=1 \leftrightarrow N=2$
    induced by  $d^\dagger_{\alpha\uparrow}$ and $d^\dagger_{\alpha\downarrow}$, $\alpha=1,2$.
     The notation $\vert 2^\prime (s_z) \rangle$, with $s_z=0,\pm 1$, specifies which one of the triplet elements is addressed. }
    \label{tab:12}
\end{table}

\vspace{1cm}

\begin{table}[h]
    \centering
    \tiny\renewcommand{\arraystretch}{1.5}

    \begin{tabular}{|c|ccccc|}
            \hline
            &&&$2 \leftrightarrow 3$&&\\
            \hline\hline
            &&$\vert{3o \uparrow}\rangle$&$\vert{3o \downarrow}\rangle$&$\vert{3e \uparrow}\rangle$&
            $\vert{3e \downarrow}\rangle$\\
            \hline
            &$\bra{2g}$&$\frac{\alpha_0+\beta_0}{2}$&$0$&$\frac{\alpha_0-\beta_0}{2}$&$0$\\
            \cline{2-6}
            &$\bra{2^\prime(+1)}$&$0$&$0$&$0$&$0$\\
          \cline{2-6}
            $d_{1\uparrow}:$&$\bra{2^\prime(0)}$&$-\frac{1}{2}$&$0$&$-\frac{1}{2}$&$0$\\
            \cline{2-6}
            &$\bra{2^\prime(-1)}$&$0$&$\frac{1}{\sqrt{2}}$&$0$&$\frac{1}{\sqrt{2}}$\\
          \cline{2-6}
            &$\bra{2^{\prime\prime}}$&$-\frac{1}{2}$&$0$&$\frac{1}{2}$&$0$\\
         \cline{2-6}
 &$\bra{2^{\prime\prime\prime}}$&$\frac{-\alpha_0+\beta_0}{2}$&$0$&$\frac{\alpha_0+\beta_0}{2}$&$0$\\

      \hline\hline

            &$\bra{2g}$&$0$&$\frac{-\alpha_0-\beta_0}{2}$&$0$&$\frac{-\alpha_0+\beta_0}{2}$\\
            \cline{2-6}
            &$\bra{2^\prime(+1)}$&$\frac{1}{\sqrt{2}}$&$0$&$\frac{1}{\sqrt{2}}$&$0$\\
          \cline{2-6}
            $d_{1\downarrow}:$&$\bra{2^\prime(0)}$&$0$&$-\frac{1}{2}$&$0$&$-\frac{1}{2}$\\
            \cline{2-6}
            &$\bra{2^\prime(-1)}$&$0$&$0$&$0$&$0$\\
          \cline{2-6}
            &$\bra{2^{\prime\prime}}$&$0$&$\frac{1}{2}$&$0$&$-\frac{1}{2}$\\
         \cline{2-6}
 &$\bra{2^{\prime\prime\prime}}$&$0$&$\frac{\alpha_0-\beta_0}{2}$&$0$&$\frac{-\alpha_0-\beta_0}{2}$\\

            \hline\hline

                &$\bra{2g}$&$\frac{-\alpha_0-\beta_0}{2}$&$0$&$\frac{\alpha_0-\beta_0}{2}$&$0$\\
            \cline{2-6}
            &$\bra{2^\prime(+1)}$&$0$&$0$&$0$&$0$\\
          \cline{2-6}
            $d_{2\uparrow}:$&$\bra{2^\prime(0)}$&$-\frac{1}{2}$&$0$&$\frac{1}{2}$&$0$\\
            \cline{2-6}
            &$\bra{2^\prime(-1)}$&$0$&$\frac{1}{\sqrt{2}}$&$0$&$-\frac{1}{\sqrt{2}}$\\
          \cline{2-6}
            &$\bra{2^{\prime\prime}}$&$-\frac{1}{2}$&$0$&$-\frac{1}{2}$&$0$\\
         \cline{2-6}
 &$\bra{2^{\prime\prime\prime}}$&$\frac{\alpha_0-\beta_0}{2}$&$0$&$\frac{\alpha_0+\beta_0}{2}$&$0$\\
            \hline\hline

            &$\bra{2g}$&$0$&$\frac{\alpha_0+\beta_0}{2}$&$0$&$\frac{-\alpha_0+\beta_0}{2}$\\
            \cline{2-6}
            &$\bra{2^\prime(+1)}$&$\frac{1}{\sqrt{2}}$&$0$&$-\frac{1}{\sqrt{2}}$&$0$\\
          \cline{2-6}
            $d_{2\downarrow}:$&$\bra{2^\prime(0)}$&$0$&$-\frac{1}{2}$&$0$&$\frac{1}{2}$\\
            \cline{2-6}
            &$\bra{2^\prime(-1)}$&$0$&0&$0$&0\\
          \cline{2-6}
            &$\bra{2^{\prime\prime}}$&$0$&$\frac{1}{2}$&$0$&$\frac{1}{2}$\\
         \cline{2-6}
 &$\bra{2^{\prime\prime\prime}}$&$0$&$\frac{-\alpha_0+\beta_0}{2}$&$0$&$\frac{-\alpha_0-\beta_0}{2}$\\

      \hline
            \end{tabular}
\caption{Matrix elements for the transition $N=2\leftrightarrow
N=3$ governed by $d_{\alpha\uparrow}$ and $d_{\alpha\downarrow}$,
$\alpha=1,2$. }
    \label{tab:32}


\vspace{1cm}


  \tiny\renewcommand{\arraystretch}{1.5}

        \begin{tabular}{|ccccc|}
            \hline
            &&$3 \leftrightarrow 4$&&\\
            \hline\hline
            &$\vert{3o \uparrow}\rangle$&$\vert{3o \downarrow}\rangle$&$\vert{3e \uparrow}\rangle$
            &$\vert{3e \downarrow}\rangle$\\
            \hline
            $d^\dagger_{1\uparrow}:\bra{2,2}$&$0$&$\frac{1}{\sqrt{2}}$&$0$&$-\frac{1}{\sqrt{2}}$\\
            \hline
            $d^\dagger_{1\downarrow}:\bra{2,2}$&$\frac{1}{\sqrt{2}}$&$0$&$-\frac{1}{\sqrt{2}}$&$0$\\
            \hline
            $d^\dagger_{2\uparrow}:\bra{2,2}$&$0$&$-\frac{1}{\sqrt{2}}$&$0$&$-\frac{1}{\sqrt{2}}$\\
            \hline
            $d^\dagger_{2\downarrow}:\bra{2,2}$&$-\frac{1}{\sqrt{2}}$&$0$&$-\frac{1}{\sqrt{2}}$&$0$\\
            \hline
        \end{tabular}
    \caption{Matrix elements for the transition $N=3 \leftrightarrow N=4$  induced by the operators
     $d^\dagger_{\alpha\uparrow}$ and $d^\dagger_{\alpha\downarrow}$, $\alpha=1,2$. }
    \label{tab:34}
\end{table}

\clearpage

\section{The master equation for the RDM in the linear
regime \label{appendix1}} We report here explicitly the coupled
equations of motion for the elements $\dot{{\rho}}_{ nm}^{(N)}(t)$
and $\dot{{\rho}}_{ nm}^{(N+1)}(t)$ of the  RDM to be solved in
the low bias regime. They are obtained from the generalized master
equation (\ref{master}) upon observing that i) in the linear
regime terms that couple states with particle numbers unlike $N$
and $N+1$ can be neglected; ii) we can reduce the sum over $h$ and
$h^\prime $
and over $l,l^\prime$
only to energy ground states. In the remaining not energy
conserving terms the sum has to go also over excited states. With
 $ \mu_{N+1}:=E^{(0)}_{N+1}-E^{(0)}_N $ being the chemical
potential we finally arrive at the  two master equations
\begin{widetext}
\begin{eqnarray}\label{Gl N}
\dot{{\rho}}_{ nm}^{(N)}(t) &=&\\\nonumber
&&\hspace{-2,5cm}-\frac{\pi}{\hbar}\sum_{\alpha=s,d}\vert{t^{\alpha}}\vert^{2}\sum_{\sigma_{\odot},\sigma_{\odot}^{\prime}}
\quad\left\lbrace
\sum_{l\in{\vert{N-1}\rangle}},\quad\sum_{j\in\vert
E^{(0)}_N,N\rangle},\quad\sum_{h,h^{\prime}\in{\vert{E^{(0)}_{N+1},N+1}\rangle}},
\quad\sum_{\widehat{h}\in{\vert{N+1}\rangle}}\right\rbrace
\left\lbrace \right. \\\nonumber
&&\hspace{-4.0cm}\qquad+\Phi_{\alpha\sigma_\odot\sigma_\odot^\prime}F^{+}_{\alpha\sigma_\odot\sigma_\odot^{\prime}}
(\mu_{N+1})\quad\big(d_{\alpha\sigma_\odot}\big)_{nh}\quad\big(d^{\dagger}_{\alpha\sigma_\odot^\prime}\big)_{hj}
\quad{\rho}_{ jm}^{(N)}(t)\\\nonumber
&&\hspace{-4.0cm}\qquad+\Phi_{\alpha\sigma_\odot\sigma_\odot^\prime}\frac{i}{\pi}P^{+}_{\alpha\sigma_\odot\sigma_\odot^{\prime}}
(\varepsilon_{\widehat{h}}-\varepsilon_{j})\quad\big(d_{\alpha\sigma_\odot}\big)_{n\widehat{h}}\quad
\big(d^{\dagger}_{\alpha\sigma_\odot^\prime}\big)_{\widehat{h}j}\quad{\rho}_{
jm}^{(N)}(t)\\\nonumber
&&\hspace{-3.9cm}\qquad-\Phi^\ast_{\alpha\sigma_\odot\sigma_\odot^\prime}\frac{i}{\pi}\left[
P^{-}_{\alpha\sigma_\odot\sigma_\odot^{\prime}}(\varepsilon_{j}-\varepsilon_{l})+R_{\alpha\sigma_\odot\sigma^{\prime}_\odot}\right]
\big(d^{\dagger}_{\alpha\sigma_\odot}\big)_{nl}\big(d_{\alpha\sigma_\odot^\prime}\big)_{lj}{\rho}_{
jm}^{(N)}(t)\\\nonumber
&&\hspace{-4.0cm}\qquad+\Phi_{\alpha\sigma_\odot\sigma_\odot^\prime}F^{+}_{\alpha\sigma_\odot\sigma_\odot^{\prime}}
(\mu_{N+1})\quad{\rho}_{ nj}^{(N)}
(t)\quad\big(d_{\alpha\sigma_\odot}\big)_{jh}\quad\big(d^{\dagger}_{\alpha\sigma_\odot^\prime}\big)_{hm}\\\nonumber
&&\hspace{-4.0cm}\qquad-\Phi_{\alpha\sigma_\odot\sigma_\odot^\prime}\frac{i}{\pi}P^{+}_{\alpha\sigma_\odot\sigma_\odot^{\prime}}
(\varepsilon_{\widehat{h}}-\varepsilon_{j})\quad{\rho}_{ nj}^{(N)}
(t)\quad\big(d_{\alpha\sigma_\odot}\big)_{j\widehat{h}}
\quad\big(d^{\dagger}_{\alpha\sigma_\odot^\prime}\big)_{\widehat{h}m}\\\nonumber
&&\hspace{-4.0cm}\qquad+\Phi^\ast_{\alpha\sigma_\odot\sigma_\odot^\prime}\frac{i}{\pi}\left[
P^{-}_{\alpha\sigma_\odot\sigma_\odot^{\prime}}(\varepsilon_{j}-\varepsilon_{l})+R_{\alpha\sigma_\odot\sigma^{\prime}_\odot}\right]
{\rho}_{ nj}^{(N)}
(t)\big(d^{\dagger}_{\alpha\sigma_\odot}\big)_{jl}\big(d_{\alpha\sigma_\odot^\prime}\big)_{lm}\\\nonumber
&&\hspace{-4.0cm}\qquad-2\Phi_{\alpha\sigma_\odot\sigma_\odot^\prime}F^{-}_{\alpha\sigma_\odot\sigma_\odot^{\prime}}
(\mu_{N+1})\quad\big(d_{\alpha\sigma_\odot}\big)_{nh^\prime}\quad{\rho}_{
h^{\prime}h}^{(N+1)}(t)
\quad\big(d^{\dagger}_{\alpha\sigma_\odot^\prime}\big)_{hm}\left. \right\rbrace ,
\end{eqnarray}
\begin{eqnarray} \label{Gl N+1}
\dot{\rho}_{ nm}^{(N+1)}(t) &=&\\\nonumber
&&\hspace{-2,5cm}-\frac{\pi}{\hbar}\sum_{\alpha=s,d}\vert{t^{\alpha}}\vert^{2}\sum_{\sigma_{\odot},\sigma_{\odot}^{\prime}}\quad\left\lbrace
\sum_{\widehat{l}\in{\vert{E_N ,N}\rangle}},
\quad\sum_{l,l^\prime\in{\vert{E^{(0)}_N,
N}\rangle}},\quad\sum_{j\in\vert{E^{(0)}_{N+1},N+1}\rangle},\quad\sum_{h\in{\vert{E_{N+2},N+2}\rangle}}\right\rbrace
\left\lbrace \right. \\\nonumber
&&\hspace{-4.0cm}\qquad+\Phi_{\alpha\sigma_\odot\sigma_\odot^\prime}\frac{i}{\pi}
P^{+}_{\alpha\sigma_\odot\sigma_\odot^{\prime}}(\varepsilon_{h}-\varepsilon_{j})
\quad\big(d_{\alpha\sigma_\odot}\big)_{nh}\quad\big(d^{\dagger}_{\alpha\sigma_\odot^\prime}\big)_{hj}\quad{\rho}_{
jm}^{(N+1)}(t)\\\nonumber
&&\hspace{-4.0cm}\qquad+\Phi^\ast_{\alpha\sigma_\odot\sigma_\odot^\prime}F^{-}_{\alpha\sigma_\odot\sigma_\odot^{\prime}}(\mu_{N+1})
\big(d^{\dagger}_{\alpha\sigma_\odot}\big)_{nl}\big(d_{\alpha\sigma_\odot^\prime}\big)_{lj}{\rho}_{
jm}^{(N+1)}(t)\\\nonumber
&&\hspace{-4.0cm}\qquad-\Phi^\ast_{\alpha\sigma_\odot\sigma_\odot^\prime}\frac{i}{\pi}
\left[P^{-}_{\alpha\sigma_\odot\sigma_\odot^{\prime}}(\varepsilon_{j}-\varepsilon_{\widehat{l}})
+R_{\alpha\sigma_\odot\sigma^{\prime}_\odot}\right]\big(d^{\dagger}_{\alpha\sigma_\odot}\big)_{n\widehat{l}}
\big(d_{\alpha\sigma_\odot^\prime}\big)_{\widehat{l}j}{\rho}_{
jm}^{(N+1)}(t)\nonumber
\\\nonumber
&&\hspace{-4.0cm}\qquad-\Phi_{\alpha\sigma_\odot\sigma_\odot^\prime}\frac{i}{\pi}
P^{+}_{\alpha\sigma_\odot\sigma_\odot^{\prime}}(\varepsilon_{h}-\varepsilon_{j})\quad{\rho}_{
nj}^{(N+1)}(t)
\quad\big(d_{\alpha\sigma_\odot}\big)_{jh}\quad\big(d^{\dagger}_{\alpha\sigma_\odot^\prime}\big)_{hm}\\\nonumber
&&\hspace{-4.0cm}\qquad+\Phi^\ast_{\alpha\sigma_\odot\sigma_\odot^\prime}\left[F^{-}_{\alpha\sigma_\odot\sigma_\odot^{\prime}}(\mu_{N+1})
\right]{\rho}_{
nj}^{(N+1)}(t)\big(d^{\dagger}_{\alpha\sigma_\odot}\big)_{j\widehat{l}}\big(d_{\alpha\sigma_\odot^\prime}\big)_{\widehat{l}m}\\\nonumber
&&\hspace{-4.0cm}\qquad+\Phi^\ast_{\alpha\sigma_\odot\sigma_\odot^\prime}\frac{i}{\pi}\left[
P^{-}_{\alpha\sigma_\odot\sigma_\odot^{\prime}}(\varepsilon_{j}-\varepsilon_{\widehat{l}})+R_{\alpha\sigma_\odot\sigma^{\prime}_\odot}\right]{\rho}_{
nj}^{(N+1)}(t)\big(d^{\dagger}_{\alpha\sigma_\odot}\big)_{j\widehat{l}}\big(d_{\alpha\sigma_\odot^\prime}\big)_{\widehat{l}m}\\\nonumber
&&\hspace{-4.0cm}\qquad-2\Phi^\ast_{\alpha\sigma_\odot\sigma_\odot^\prime}
F^{+}_{\alpha\sigma_\odot\sigma_\odot^{\prime}}(\mu_{N+1})
\quad\big(d^\dagger_{\alpha\sigma_\odot}\big)_{nl^\prime}\quad{\rho}_{
l^{\prime}l}^{(N)}(t)
\quad\big(d_{\alpha\sigma_\odot^\prime}\big)_{lm}\left. \right\rbrace . \\
\end{eqnarray}
\end{widetext}
Notice that we kept the sums over excited states $l,\widehat{h}$
in (\ref{Gl N}) and $\widehat{l},h$ in (\ref{Gl N+1}) which are
responsible for the virtual transitions.



\end{document}